%
%
\magnification \magstep1
\hsize=16truecm
\vsize=22truecm
\voffset=1.0truecm
\hoffset=-0.1truecm
\normalbaselineskip=5.25mm
\baselineskip=5.25mm
\parskip=5pt
\parindent=20pt
\nopagenumbers
\headline={\ifnum\pageno>1 {\hss\tenrm-\ \folio\ -\hss} \else {\hfill}\fi}
\def\em{\sl}
%
%
\newcount\EQNcount \EQNcount=1
\newcount\CLAIMcount \CLAIMcount=1
\newcount\SECTIONcount \SECTIONcount=0
\newcount\SUBSECTIONcount \SUBSECTIONcount=0
\def\actualnumber{\number\SECTIONcount}
\newcount\timecount
\def\TODAY{\number\day~\ifcase\month\or January \or February \or March \or
  April \or May \or June
  \or July \or August \or September \or October \or November \or December \fi
  \number\year\timecount=\number\time
  \divide\timecount by 60}
\newdimen\strutdepth
\def\DRAFT{\def\lmargin(##1){\strut\vadjust{\kern-\strutdepth
  \vtop to \strutdepth{
  \baselineskip\strutdepth\vss\rlap{\kern-1.2 truecm\eightpoint{##1}}}}}
  \font\footfont=cmti7
  \footline={{\footfont \hfil File:\jobname, \TODAY,  \number\timecount h}}}
\def\lmargin(#1){}
\def\ifundefined#1{\expandafter\ifx\csname#1\endcsname\relax}
\def\ifff(#1,#2,#3){\ifundefined{#1#2}
  \expandafter\xdef\csname #1#2\endcsname{#3}\else
  \write16{Warning : doubly defined #1,#2}\fi}
\def\NEWDEF #1,#2,#3 {\ifff({#1},{#2},{#3})}
\def\EQ(#1){\lmargin(#1)\eqno\tag(#1)}
\def\NR(#1){&\lmargin(#1)\tag(#1)\cr}  
\def\tag(#1){({\rm \actualnumber.\number\EQNcount})
  \NEWDEF e,#1,(\actualnumber.\number\EQNcount)
  \global\advance\EQNcount by 1
  }
\def\equ(#1){\ifundefined{e#1}$\spadesuit$#1\else\csname e#1\endcsname\fi}
\def\CLAIM #1(#2) #3\par{
  \vskip.1in\medbreak\noindent
  {\lmargin(#2)\bf #1~\actualnumber.\number\CLAIMcount.} {\sl #3}\par
  \NEWDEF c,#2,{#1~\actualnumber.\number\CLAIMcount}
  \global\advance\CLAIMcount by 1
  \ifdim\lastskip<\medskipamount
  \removelastskip\penalty55\medskip\fi}
\def\CLAIMNONR #1 #2\par{
  \vskip.1in\medbreak\noindent
  {\bf #1.} {\sl #2}\par
  \ifdim\lastskip<\medskipamount
  \removelastskip\penalty55\medskip\fi}
\def\clm(#1){\ifundefined{c#1}$\spadesuit$#1\else\csname c#1\endcsname\fi}
\def\sectionsize{\twelvepoint}
\def\sectiontype{\bf}
\newskip\beforesectionskipamount  
\newskip\sectionskipamount        
\beforesectionskipamount=24pt plus8pt minus8pt
\sectionskipamount=3pt plus1pt minus1pt
\def\sectionskip{\vskip\sectionskipamount}
\def\beforesectionskip{\vskip\beforesectionskipamount}
\def\SECTION#1\par{\vskip0pt plus.3\vsize\penalty-75
  \vskip0pt plus -.3\vsize
  \global\advance\SECTIONcount by 1
  \def\actualnumber{\number\SECTIONcount}
  \beforesectionskip\noindent
  {\sectionsize\sectiontype \actualnumber.\ #1}
  \EQNcount=1
  \CLAIMcount=1
  \SUBSECTIONcount=0
  \nobreak\sectionskip\noindent}
\def\SECTIONNONR#1\par{\vskip0pt plus.3\vsize\penalty-75
  \vskip0pt plus -.3\vsize
  \global\advance\SECTIONcount by 1
  \beforesectionskip\noindent
  {\sectionsize\sectiontype #1}
  \EQNcount=1
  \CLAIMcount=1
  \SUBSECTIONcount=0
  \nobreak\sectionskip\noindent}
\def\SECT(#1)#2\par{\lmargin(#1)
  \SECTION #2\par
  \NEWDEF s,#1,{\actualnumber}}
\def\sec(#1){\ifundefined{s#1}$\spadesuit$#1
  \else Section \csname s#1\endcsname\fi}
\def\subsectionsize{}
\def\subsectiontype{\bf}
\def\SUBSECTION#1\par{\vskip0pt plus.2\vsize\penalty-75
  \vskip0pt plus -.2\vsize
  \global\advance\SUBSECTIONcount by 1
  \beforesectionskip\noindent
  {\subsectionsize\subsectiontype \actualnumber.\number\SUBSECTIONcount.\ #1}
  \nobreak\sectionskip\noindent}
\def\SUBSECT(#1)#2\par{\lmargin(#1)
  \SUBSECTION #2\par
  \NEWDEF p,#1,{\actualnumber.\number\SUBSECTIONcount}}
\def\subsec(#1){\ifundefined{p#1}$\spadesuit$#1
  \else Subsection \csname p#1\endcsname\fi}
\def\APPENDIX(#1)#2\par{
  \def\actualnumber{#1}
  \SECTIONNONR Appendix #1. #2\par}
\def\REFERENCES{
  \parindent=30pt
  \parskip=5pt  
  \SECTIONNONR References\par}

\def\PROOF{\medskip\noindent{\bf Proof.\ }}
\def\REMARK{\medskip\noindent{\bf Remark.\ }}
\def\REMARKS{\medskip\noindent{\bf Remarks.\ }}

\def\LIKEREMARK(#1){\medskip\noindent{\bf #1.\ }}
%
%
\let\endarg=\par
\def\finish{\def\endarg{\par\endgroup}}
\def\start{\endarg\begingroup}
\def\getNORMAL#1{{#1}}
\def\titlesize{\twelvepoint}
\def\titletype{\bf}
\def\TITLE{\beginTITLE\getTITLE}
  \def\beginTITLE{\start
     \titlesize\titletype\baselineskip=1.728
     \normalbaselineskip\rightskip=0pt plus1fil
     \noindent
     \def\endarg{\par\vskip.35in\endgroup}}
  \def\getTITLE{\getNORMAL}
\def\ENDTITLE{\endarg}
\def\AUTHOR{\beginAUTHOR\getAUTHOR}
  \def\beginAUTHOR{\start
    \vskip .25in\rm\noindent\finish}
  \def\getAUTHOR{\getNORMAL}
\def\FROM{\beginFROM\getFROM}
  \def\beginFROM{\start\parskip=0pt\vskip\baselineskip
    \def\finish{\def\endarg{\egroup\par\endgroup}}
    \vbox\bgroup\obeylines\eightpoint\em\finish}
  \def\getFROM{\getNORMAL}
\def\ABSTRACT#1\par{
  \vskip 1in {\noindent\sectionsize\sectiontype Abstract.} #1 \par}
\def\ENDABSTRACT{\vfill\break}
%
%
\newdimen\texpscorrection
\texpscorrection=0truecm  
\newcount\FIGUREcount \FIGUREcount=0
\newskip\ttglue 
\newdimen\figcenter
\def\figure #1 #2 #3 #4\cr{\null
  \global\advance\FIGUREcount by 1
  \NEWDEF fig,#1,{Fig.~\number\FIGUREcount}
  \write16{ FIG \number\FIGUREcount: #1}
  {\goodbreak\figcenter=\hsize\relax
  \advance\figcenter by -#3truecm
  \divide\figcenter by 2
  \midinsert\vskip #2truecm\noindent\hskip\figcenter
  \includegraphics{#1}
  \vskip 0.8truecm\noindent \vbox{\eightpoint\noindent
  {\bf\fig(#1)}: #4}\endinsert}}
\def\figurewithtex #1 #2 #3 #4 #5\cr{\null
  \global\advance\FIGUREcount by 1
  \NEWDEF fig,#1,{Fig.~\number\FIGUREcount}
  \write16{ FIG \number\FIGUREcount: #1}
  {\goodbreak\figcenter=\hsize\relax
  \advance\figcenter by -#4truecm
  \divide\figcenter by 2
  \midinsert\vskip #3truecm\noindent\hskip\figcenter
  \includegraphics{#1}{\hskip\texpscorrection\input #2 }
  \vskip 0.8truecm\noindent \vbox{\eightpoint\noindent
  {\bf\fig(#1)}: #5}\endinsert}}
\def\fig(#1){\ifundefined{fig#1}$\spadesuit$#1
  \else\csname fig#1\endcsname\fi}
%
%
\catcode`@=11
\def\footnote#1{\let\@sf\empty 
  \ifhmode\edef\@sf{\spacefactor\the\spacefactor}\/\fi
  #1\@sf\vfootnote{#1}}
\def\vfootnote#1{\insert\footins\bgroup\eightpoint
  \interlinepenalty\interfootnotelinepenalty
  \splittopskip\ht\strutbox 
  \splitmaxdepth\dp\strutbox \floatingpenalty\@MM
  \leftskip\z@skip \rightskip\z@skip \spaceskip\z@skip \xspaceskip\z@skip
  \textindent{#1}\footstrut\futurelet\next\fo@t}
\def\fo@t{\ifcat\bgroup\noexpand\next \let\next\f@@t
  \else\let\next\f@t\fi \next}
\def\f@@t{\bgroup\aftergroup\@foot\let\next}
\def\f@t#1{#1\@foot}
\def\@foot{\strut\egroup}
\def\footstrut{\vbox to\splittopskip{}}
\skip\footins=\bigskipamount 
\count\footins=1000 
\dimen\footins=8in  
\catcode`@=12       
%
\font\twelverm=cmr12
\font\twelvei=cmmi12
\font\twelvesy=cmsy10 scaled\magstep1
\font\twelveex=cmex10 scaled\magstep1
\font\twelvebf=cmbx12 
\font\twelvett=cmtt12
\font\twelvesl=cmsl12
\font\twelveit=cmti12
\font\ninerm=cmr9

\font\ninesy=cmsy9

\font\eightrm=cmr8
\font\eighti=cmmi8
\font\eightsy=cmsy8
\font\eightex=cmex8
\font\eightbf=cmbx8
\font\eighttt=cmtt8
\font\eightsl=cmsl8
\font\eightit=cmti8
\font\sixrm=cmr6
\font\sixi=cmmi6
\font\sixsy=cmsy6
\font\sixbf=cmbx6
\newfam\truecmr 
\newfam\truecmsy
\font\twelvetruecmr=cmr10 scaled\magstep1
\font\twelvetruecmsy=cmsy10 scaled\magstep1
\font\tentruecmr=cmr10
\font\tentruecmsy=cmsy10
\font\eighttruecmr=cmr8
\font\eighttruecmsy=cmsy8
\font\seventruecmr=cmr7
\font\seventruecmsy=cmsy7
\font\sixtruecmr=cmr6
\font\sixtruecmsy=cmsy6
\font\fivetruecmr=cmr5
\font\fivetruecmsy=cmsy5
\textfont\truecmr=\tentruecmr
\scriptfont\truecmr=\seventruecmr
\scriptscriptfont\truecmr=\fivetruecmr
\textfont\truecmsy=\tentruecmsy
\scriptfont\truecmsy=\seventruecmsy
\scriptscriptfont\truecmsy=\fivetruecmsy
%
\def \eightpoint{\def\rm{\fam0\eightrm}
  \textfont0=\eightrm \scriptfont0=\sixrm \scriptscriptfont0=\fiverm 
  \textfont1=\eighti  \scriptfont1=\sixi  \scriptscriptfont1=\fivei 
  \textfont2=\eightsy \scriptfont2=\sixsy \scriptscriptfont2=\fivesy 
  \textfont3=\eightex \scriptfont3=\eightex \scriptscriptfont3=\eightex
  \textfont\itfam=\eightit          \def\it{\fam\itfam\eightit}%
  \textfont\slfam=\eightsl          \def\sl{\fam\slfam\eightsl}%
  \textfont\ttfam=\eighttt          \def\tt{\fam\ttfam\eighttt}%
  \textfont\bffam=\eightbf          \scriptfont\bffam=\sixbf
  \scriptscriptfont\bffam=\fivebf   \def\bf{\fam\bffam\eightbf}%
  \textfont\truecmr=\eighttruecmr   \scriptfont\truecmr=\sixtruecmr
  \scriptscriptfont\truecmr=\fivetruecmr
  \textfont\truecmsy=\eighttruecmsy \scriptfont\truecmsy=\sixtruecmsy
  \scriptscriptfont\truecmsy=\fivetruecmsy
  \tt \ttglue=.5em plus.25em minus.15em 
  \setbox\strutbox=\hbox{\vrule height7pt depth2pt width0pt}%
  \normalbaselineskip=9pt
  \let\sc=\sixrm  \let\big=\eightbig  \normalbaselines\rm
}
\def \twelvepoint{\def\rm{\fam0\twelverm}
\textfont0=\twelverm  \scriptfont0=\tenrm  \scriptscriptfont0=\eightrm 
\textfont1=\twelvei   \scriptfont1=\teni   \scriptscriptfont1=\eighti 
\textfont2=\twelvesy  \scriptfont2=\tensy  \scriptscriptfont2=\eightsy 
\textfont3=\twelveex  \scriptfont3=\tenex  \scriptscriptfont3=\eightex 
\textfont\itfam=\twelveit             \def\it{\fam\itfam\twelveit}%
\textfont\slfam=\twelvesl             \def\sl{\fam\slfam\twelvesl}%
\textfont\ttfam=\twelvett             \def\tt{\fam\ttfam\twelvett}%
\textfont\bffam=\twelvebf             \scriptfont\bffam=\tenbf
\scriptscriptfont\bffam=\eightbf      \def\bf{\fam\bffam\twelvebf}%
\textfont\truecmr=\twelvetruecmr      \scriptfont\truecmr=\tentruecmr
\scriptscriptfont\truecmr=\eighttruecmr
\textfont\truecmsy=\twelvetruecmsy    \scriptfont\truecmsy=\tentruecmsy
\scriptscriptfont\truecmsy=\eighttruecmsy
\tt \ttglue=.5em plus.25em minus.15em 
\setbox\strutbox=\hbox{\vrule htwelve7pt depth2pt width0pt}%
\normalbaselineskip=12pt
\let\sc=\tenrm  \let\big=\twelvebig  \normalbaselines\rm
}
\catcode`@=11
\def\eightbig#1{{\hbox{$\textfont0=\ninerm\textfont2=\ninesy\left#1
  \vbox to6.5pt{}\right.\n@space$}}}
\catcode`@=12
%

\def\CC{{\cal C}}

\def\HH{{\cal H}}

\def\NN{{\cal N}}
\def\OO{{\cal O}}

\def\QQ{{\cal Q}}
\def\RR{{\cal R}}

\def\VV{{\cal V}}

\def\HB {\hfill\break}
\def\sqr#1#2{{\vcenter{\vbox{\hrule height .#2pt
        \hbox{\vrule width.#2pt height#1pt \kern#1pt
           \vrule width.#2pt}
        \hrule height.#2pt}}}}
\def\square{\mathchoice\sqr64\sqr64\sqr{2.1}3\sqr{1.5}3}        
\def\QED{\hfill$\square$}

\def\L{{\rm L}}
\def\H{{\rm H}}
\def\DS{\displaystyle}
\mathchardef\pr=`'    
\def\prp{\pr\!\!{}_+}
\def\pra{\pr\!\!\!{}_3}
\def\prb{\pr\!\!\!{}_4}
\def\pap{\rm }
\def\bok{\sl }
\def\real{\relax\ifmmode {{\rm I} \! {\rm R}}                           
 \else${{\rm I} \! {\rm R}}$\fi}                                        
\def\integer{\relax\ifmmode {\!\rlap{Z}\hskip 4.2pt{\rm Z}}             
 \else${\!\rlap{Z}\hskip 4pt{\rm Z}}$\fi}                               
\def\natural{\relax\ifmmode {{\rm I} \! {\rm N}}                        
 \else${{\rm I} \! {\rm N}} $\fi}                                       
\def\complex{\relax\ifmmode {\!\rlap{\rm C}\hskip 2.9pt                 
 {\vrule height 8 pt width 0.5 pt}\,\,}                                 
 \else${\!\rlap{\rm C}\hskip 2.9pt                                      
 {\vrule height 8 pt width 0.5pt}\,\,}$\fi}                             
\hfuzz=1pt

\TITLE Existence and Stability of Propagating Fronts for an 
       Autocatalytic Reaction-Diffusion System
\ENDTITLE
\AUTHOR S. Focant

\FROM Physique Th\'eorique
      Universit\'e Catholique de Louvain
      2, ch. du cyclotron
      B-1348 Louvain-la-Neuve, Belgique

\AUTHOR Th. Gallay

\FROM Analyse Num\'erique et EDP
      Universit\'e de Paris-Sud
      B\^atiment 425
      F-91405 Orsay, France
\ENDTITLE         

\ABSTRACT
We study a one-dimensional reaction-diffusion system which describes an 
isothermal autocatalytic chemical reaction involving both a quadratic 
($A + B \to 2B$) and a cubic ($A + 2B \to 3B$) autocatalysis. The parameters 
of this system are the ratio $D = D_B/D_A$ of the diffusion constants of the 
reactant $A$ and the autocatalyst $B$, and the relative activity $k$ of the 
cubic reaction. First, for all values of $D > 0$ and $k \ge 0$, we prove the 
existence of a family of propagating fronts (or travelling waves) describing 
the advance of the reaction.  In particular, in the quadratic case $k=0$, we 
recover the results of Billingham and Needham [BN]. Then, if $D$ is close to 
$1$ and $k$ is sufficiently small, we prove using energy functionals that 
these propagating fronts are stable against small perturbations in 
exponentially weighted Sobolev spaces. This extends to our system part of 
the stability results which are known for the scalar Fisher equation.
\ENDABSTRACT

\SECTION Introduction

We consider the reaction-diffusion system
$$ \eqalign{
   \partial_t u(x,t) \,&=\, \phantom{D}\partial_x^2 u(x,t) - u(x,t)v(x,t) - 
      k u(x,t) v(x,t)^2 ~, \cr
   \partial_t v(x,t) \,&=\, D \partial_x^2 v(x,t) + u(x,t)v(x,t) + 
      k u(x,t) v(x,t)^2 ~,} \EQ(rd)
$$
where $u,v$ are nonnegative functions of $(x,t) \in \real \times \real_+$,
and $D > 0$, $k \ge 0$ are constant parameters. This system describes
(in dimensionless variables) an isothermal autocatalytic chemical reaction
of mixed order, involving both a quadratic ($A + B \to 2B$) and a cubic 
($A + 2B \to 3B$) autocatalysis, see [HPSS], [BN]. Here, $u$ and $v$ are
the concentrations of the reactant $A$ and the autocatalyst $B$, and 
$D = D_B/D_A$ is the ratio of the diffusion constants. The parameter $k$ 
measures the contribution of the cubic autocatalysis to the whole reaction. 
Of particular interest are the purely quadratic case $k=0$, and the purely 
cubic case ``$k=+\infty$'' which corresponds, after a rescaling, to 
Eq.\equ(rd) with $k=1$ and without quadratic terms.  

The dynamics of the system \equ(rd) on a bounded domain $\Omega \subset \real$ 
with homogeneous boundary conditions is well understood [Ma], [HY]. 
For any initial data $u_0,v_0 \in \L^\infty(\Omega)$, the solution
$(u(t),v(t)) \equiv (u(\cdot,t),v(\cdot,t)) \in \L^\infty(\Omega)^2$ is 
uniformly bounded for all times and converges as $t \to +\infty$ to a 
uniform steady state $(u^*,v^*)$ satisfying $u^* v^* = 0$. On the other hand, 
if $u_0,v_0 \in \L^1(\real) \cap \L^\infty(\real)$, the solution ($u(t),
v(t)$) of the system \equ(rd) on the whole real line stays 
bounded for all times [BKX] and converges uniformly to zero as 
$t \to +\infty$. In the purely cubic case, a very detailed description 
of this convergence can be found in [BX], [BKX].
Finally, if $u_0,v_0 \in \L^\infty(\real)$ only, then the solution
$(u(t),v(t))$ exists for all times, and stays uniformly bounded if 
$D \le 1$ [MP]; if $D > 1$, uniform boundedness is an open problem, but 
an upper bound is known which diverges extremely slowly as 
$t \to +\infty$ [CX]. Of course, very little is known about the 
behavior of the solutions in this general situation. 

In this paper, we investigate the existence and stability of propagating
fronts (or travelling waves) for the system \equ(rd). These are 
uniformly translating solutions connecting the stable steady state
$(u,v) = (0,1)$ at $x = -\infty$ to the unstable state $(u,v) = (1,0)$
at $x = +\infty$. Thus we look for solutions of \equ(rd) of the form
$u(x,t) = \alpha(x-ct)$, $v(x,t) = \beta(x-ct)$, where $c > 0$ is 
the velocity of the front. The nonnegative functions $\alpha,\beta$
satisfy the system
$$ \eqalign{
   \alpha''(x) + c \alpha'(x) - \alpha \beta - k \alpha \beta^2 \,&=\, 0~,\cr
   D\beta''(x) + c \beta'(x) + \alpha \beta + k \alpha \beta^2 \,&=\, 0~,}
   \EQ(front)
$$
together with the boundary conditions
$$
   (\alpha(-\infty), \beta(-\infty)) \,=\, (0,1)~, \quad
   (\alpha(+\infty), \beta(+\infty)) \,=\, (1,0)~. \EQ(bcond)
$$

\REMARK It is easy to verify that the nonnegative time-independent
solutions of \equ(rd) are exactly the uniform steady states $(u,v)=(a,b)$ 
with $a \ge 0$, $b \ge 0$ satisfying $ab=0$. Moreover, a necessary 
condition for the existence of a trajectory of \equ(front) connecting 
$(a_-,b_-)$ at $x = -\infty$ to $(a_+,b_+)$ at $x =+\infty$ is that 
$a_- + b_- = a_+ + b_+$. Indeed, adding the two equations in 
\equ(front) and integrating with respect to $x$, we obtain the conservation
law
$$
   \alpha'(x) + c \alpha(x) + D \beta'(x) + c \beta(x) \,=\, {\rm const.}~,
   \EQ(claw)
$$
and the assertion follows by taking the limits $x \to \pm\infty$. 
Therefore, any heteroclinic orbit of \equ(front) must connect $(0,a)$ 
to $(a,0)$ for some $a > 0$. Now, the system \equ(rd) is invariant under 
the scaling transformation
$$
   u(x,t) \to \lambda^2 u(\lambda x,\lambda^2 t)~, \quad
   v(x,t) \to \lambda^2 v(\lambda x,\lambda^2 t)~, \quad
   k \to k/\lambda^2~, \EQ(scaling)
$$
for all $\lambda > 0$, so there is no loss of generality in assuming that
$a=1$. This explains the choice of the boundary conditions \equ(bcond). 

Existence of solutions to \equ(front), \equ(bcond) has been studied by 
Billingham and Needham [BN] for all $D > 0$ in the cases $k =0$ and 
$k=+\infty$. First, they show that any solution satisfies the following 
properties for all $x \in \real$:
$$ \matrix{
   0 < \alpha(x) < 1~, & \alpha'(x) > 0~,\cr  
   0 < \beta(x) < 1~, & \beta'(x) < 0~,} \quad \hbox{and} \quad 
   \cases{\alpha(x) + \beta(x) \le 1 & if $D \le 1$, \cr
          \alpha(x) + \beta(x) \ge 1 & if $D \ge 1$.}
   \EQ(prop)
$$
Then, in the purely quadratic case, they prove that
a travelling wave exists if and only if $c \ge 2\sqrt{D}$, and is unique 
up to a translation in the variable $x$. Finally, in the purely cubic case, 
they argue that a propagating front exists if and only if $c \ge v_2^*(D)$, 
where $v_2^*(D)$ is some increasing function of $D$ satisfying $v_2^*(1) =
1/\sqrt{2}$, $v_2^*(D) = \OO(D)$ as $D \to 0$, and $v_2^*(D) = \OO(\sqrt{D})$
as $D \to +\infty$. In this latter case, their argument relies in part 
on numerical calculations. 

On the other hand, if $D=1$, the existence of travelling waves has been 
proved for all $k \ge 0$. Indeed, it follows from \equ(prop)
that $\alpha(x) +\beta(x) =1$ in this case, so that $\beta$ satisfies the
single equation
$$
   \beta''(x) + c\beta'(x) + \beta(x)(1-\beta(x))(1+k\beta(x)) \,=\, 0~, 
   \EQ(single)
$$
together with the boundary conditions $\beta(-\infty)=1$, $\beta(+\infty)=0$. 
This problem can be studied by usual phase space techniques, and is known
to have a nonnegative solution if and only if $c \ge c^*(k)$, where
$$
   c^*(k) \,=\, \cases{ 2 & if $k \le 2$~, \cr
   \sqrt{k/2}+\sqrt{2/k} & if $k > 2$~,} \EQ(cstar)
$$
see [BBDKL], [vS]. If $k \le 2$, the front with minimal speed is often called
``pulled'' or ``linear'', because its velocity $c^*$ and its decay rate 
at infinity can be determined from the linearized equation ahead of the 
front. In the converse case, it is called ``pushed'' or ``nonlinear''. 

In the general case $D > 0$, $k \ge 0$, the situation seems more complicated,
and our results are still incomplete. If $D > 1$, we can still prove 
the existence of a minimal propagation speed for the travelling waves:

\CLAIM Theorem(Dlarge) Let $D \ge 1$, $k \ge 0$. Then there exists
$c^* = c^*(D,k) > 0$ such that Eqs.\equ(front), \equ(bcond) have a nonnegative 
solution if and only if $c \ge c^*$. This solution is unique up to a 
translation and satisfies \equ(prop). Moreover, $c^*(D,k)$ is a
non-decreasing function of $k$ and satisfies
$$
   \left. \matrix{ k \le 2 &\!:\!& 2\sqrt{D} \cr k > 2 &\!:\!& 
   \sqrt{D} (\sqrt{k/2} + \sqrt{2/k})} \right\} \,\le\, 
   c^*(D,k) \,\le\, \left\{ \matrix{2\sqrt{D} &\!:\!& k \le {3D-1 \over 3D-2} 
   \cr \bar c(D,k) &\!:\!& k > {3D-1 \over 3D-2}} \right. \EQ(bounds)
$$
where $\bar c(D,k)$ is given by Eq.(2.13) below. In particular, $\bar c(D,k)
\le \sqrt{D}(\sqrt{k}+\sqrt{1/k})$ for all $D \ge 1$ and all $k > 
(3D-1)/(3D-2)$. 

An immediate consequence of \clm(Dlarge) is:

\CLAIM Corollary(kstar) Let $D \ge 1$. There exists $k^* = k^*(D)$ such 
that the minimal speed $c^*(D,k)$ defined in \clm(Dlarge) satisfies 
$c^* = 2\sqrt{D}$ if $k < k^*(D)$ and $c^* > 2\sqrt{D}$ if $k > k^*(D)$. 
Moreover, one has
$$
   {3D-1 \over 3D-2 } \,\le\, k^*(D) \,\le\, 2~,
$$
for all $D \ge 1$. 

This result says that the front with minimal speed is ``linear'' if $k < k^*$
and ``nonlinear'' if $k > k^*$. A numerical determination of the curve 
$k^*(D)$ is shown in Fig.~1 below. 

On the other hand, if $D < 1$, we do not know whether the set of values
of $c$ for which a propagating front exists is always an interval of the 
form $[c^*,\infty)$. However, for all $D < 1$ and all 
$k \ge 0$, we do know that a nonnegative solution to Eqs.\equ(front), 
\equ(bcond) exists if $c$ is sufficiently large and does not exist if
$c < 2\sqrt{D}$:

\CLAIM Theorem(Dsmall) Let $0 < D < 1$, $k \ge 0$. Then Eqs.\equ(front),
\equ(bcond) have a nonnegative solution if
$$
   c \,\ge\, \cases{ 2\sqrt{D} & if $k \le 2$~, \cr
             \sqrt{D} (\sqrt{k/2}+\sqrt{2/k}) & if $k > 2$~. }
$$
This solution is unique up to a translation and satisfies \equ(prop).
Conversely, there exists no nonnegative solution of Eqs.\equ(front), 
\equ(bcond) if $c < 2\sqrt{D}$. 

\REMARK In particular, \clm(Dlarge) and \clm(Dsmall) show that, if 
$D \ge 1$ and $k \le (3D-1)/(3D-2)$, or if $D < 1$ and $k \le 2$, a 
nonnegative propagating front exists if and only if $c \ge 2\sqrt{D}$.
This extends the result obtained by Billingham and Needham [BN] for $k=0$.  

\figure 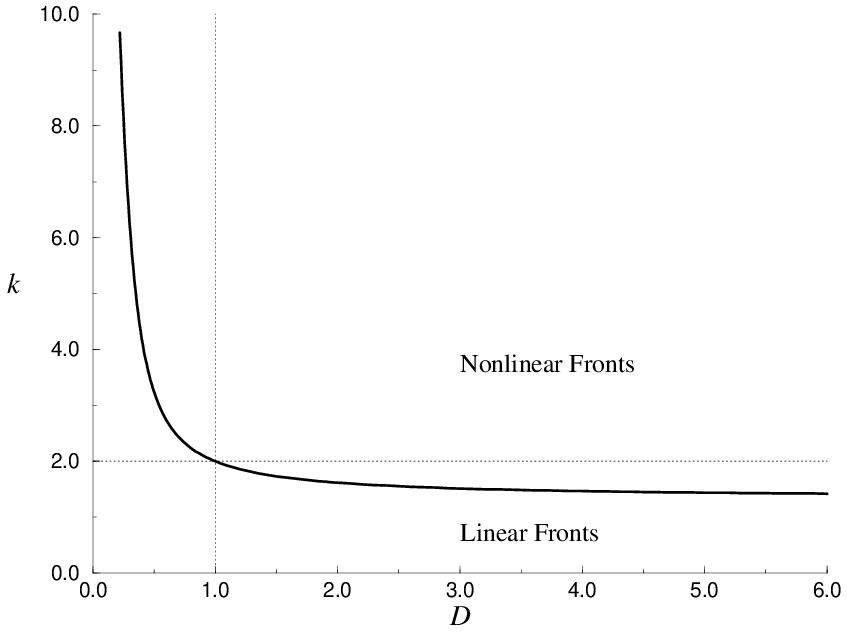 7 11 The curve $k^*(D)$ separating the parameter regions
where $c^*(D,k) = 2\sqrt{D}$ (linear fronts) and $c^*(D,k) > 2\sqrt{D}$
(nonlinear fronts). The existence of this curve is asserted by \clm(kstar)
for $D \ge 1$ only, but numerically this separatrix can be observed for all
$D > 0$. Note that $k^*(D) > 2$ if $0 < D < 1$, in agreement with 
\clm(Dsmall). \cr

It follows from \clm(Dlarge) and \clm(Dsmall) that, for all $D > 0$ and all 
$k \ge 0$, the system \equ(rd) has a one-parameter family of uniformly 
translating front solutions, indexed by the velocity $c$. A natural question 
to address is whether these propagating fronts are stable against
sufficiently small perturbations in appropriate function spaces. 
Again, the case $D = 1$ is easier and can be treated separately. 
Indeed, if $D=1$, the system \equ(rd) can be written as 
$$ \eqalign{
   \partial_t w(x,t) \,&=\, \partial_x^2 w(x,t) ~,\cr
   \partial_t v(x,t) \,&=\, \partial_x^2 v(x,t) + v(x,t)(w(x,t)-v(x,t))
   (1+kv(x,t))~,} \EQ(modif)
$$
where $w(x,t) = u(x,t) + v(x,t)$. In these new variables, the propagating
fronts are given by $w(x,t) = 1$, $v(x,t) = \beta(x-ct)$, where $\beta$ is 
the solution of \equ(single). Therefore, if we consider initial data of the
form $w_0 = 1+f$, $v_0 = \beta+g$, where $f,g$ are sufficiently localized 
perturbations, then the solution of \equ(modif) will satisfy $w(x,t) 
= 1 + \OO(t^{-1/2})$ as $t \to +\infty$, so that the behavior of $v(x,t)$ 
for large times will be governed by the nonlinear diffusion equation 
$$
   \partial_t v(x,t) \,=\, \partial_x^2 v(x,t) + v(x,t)(1-v(x,t))
   (1+kv(x,t))~, \EQ(scalar)
$$
up to a remainder which can be controlled rigorously [Fo1]. Now, the stability 
of the travelling wave solutions of Eq.\equ(scalar) has been intensively 
studied by many authors [AW], [Sa], [Ki], [EW], [BK], [Ga2]. In particular, 
for all $k \ge 0$ and all $c \ge c^*(k)$, each individual front is known to be 
asymptotically stable against perturbations which decay to 
zero sufficiently fast --- at least as fast as the front itself --- as 
$x \to +\infty$. The decay rate
in time of the perturbations is polynomial or exponential depending on 
the choice of the function space. In addition, if $k > 2$ and $c = c^*$ 
(the ``pushed'' case), the family of all translates of the front is
orbitally stable against perturbations which decay even {\sl slower} than
the front itself as $x \to +\infty$. 

In the general case $D \neq 1$, the reduction to a single equation is 
no longer possible, and much less is known about the stability of 
propagating fronts. We shall restrict ourselves in the sequel
to the situation where $D$ is close to $1$ and $k$ is close to $0$, see
the remarks after Theorem~1.4 below for a discussion of these limitations.
In this parameter region, we know from \clm(Dlarge), \clm(Dsmall) that
a propagating front $(\alpha,\beta)$ exists if and only if $c \ge 2\sqrt{D}$.
Since $\alpha(x) \to 0$ and $\beta(x) \to 1$ as $x \to -\infty$, we shall 
assume (without loss of generality) that $\beta(x) - \alpha(x) \ge 3/4$ 
for all $x \le 0$. Setting 
$$
   u(x,t) = \alpha(x-ct) + f(x-ct,t)~, \quad
   v(x,t) = \beta(x-ct)  + g(x-ct,t)~,
$$
and inserting into \equ(rd), we obtain the evolution equations for the
perturbation $(f,g)$ in the moving frame
$$ \eqalign{
   \partial_t f \,&=\, \phantom{D}\partial_x^2 f + c \partial_x f 
      - \beta(1+k\beta)f - \alpha(1+2k\beta)g - N(f,g)~, \cr
   \partial_t g \,&=\, D \partial_x^2 g + c \partial_x g + \alpha(1+2k\beta)g 
      + \beta(1+k\beta)f + N(f,g)~,} \EQ(fg)
$$
where $ N(f,g) = fg + k(2\beta fg + \alpha g^2 + fg^2)$.

As in the scalar case, it is necessary to use weighted spaces which force
the perturbations $(f,g)$ to decay to zero sufficiently fast as 
$x \to +\infty$ [Sa]. For any $s > 0$, we consider the Hilbert spaces $X_s$, 
$Y_s$ of real functions on $\real$ defined by the norms
$$
   \|h\|_{X_s}^2 \,=\, \int_\real |h(x)|^2 (1+e^{2sx}) \,dx~, \quad
   \|h\|_{Y_s}^2 \,=\, \|h\|_{X_s}^2 + \|h'\|_{X_s}^2~, \EQ(ynorm)
$$
where $'$ denotes the (space) derivative. We also note $X_s^2 = X_s \times
X_s$, $Y_s^2 = Y_s \times Y_s$. Then a direct calculation shows that 
the origin $(f,g) = (0,0)$ in \equ(fg) is linearly stable in $X_s^2$ 
only if $Ds^2 - cs + 1 \le 0$, see [Ga1], [GR] for a similar discussion. 
Therefore, we must choose $s = 1/\sqrt{D}$ if $c = 2\sqrt{D}$. If $c > 
2\sqrt{D}$, $s$ can be chosen in a whole interval, but the biggest 
perturbation space corresponds to the choice
$$
   s \,=\, {1 \over 2D} (c - \sqrt{c^2 -4D})~, \EQ(schoice)
$$
which we shall always assume in the sequel. Note that this value
corresponds to the exponential decay rate of both $\alpha(x)$, $\beta(x)$ 
as $x \to +\infty$. 

With these definitions, we can state our main result:

\CLAIM Theorem(stab) There exist $d_0 > 0$ and $k_0 > 0$ 
such that, for all $D \in [1-d_0,1+d_0]$, all $k \in [0,k_0]$ and all
$c \ge 2\sqrt{D}$, there exist $\epsilon_0 > 0$ and $K_0 \ge 1$
such that the following holds: for all $(f_0,g_0) \in Y_s^2$ satisfying
$\|(f_0,g_0)\|_{Y_s^2} \le \epsilon_0$, where $s$ is given by \equ(schoice),
Eq.\equ(fg) has a unique global solution $(f,g) \in \CC^0([0,+\infty),Y_s^2)
\cap \CC^1((0,+\infty),Y_s^2)$ satisfying $(f(0),g(0)) = (f_0,g_0)$. Moreover, 
one has
$$
   \|(f(t),g(t))\|_{Y_s^2} \,\le\, K_0 \|(f_0,g_0)\|_{Y_s^2}~, \EQ(bound)
$$
for all $t \ge 0$, and 
$$
   \lim_{t \to +\infty} \|(\partial_x f(t),\partial_x g(t))\|_{X_s^2} 
   \,=\, 0~. \EQ(limit)
$$

\REMARKS \HB
\noindent{\bf a)} In particular, Eqs.\equ(bound), \equ(limit) imply that 
the perturbation $(f(t),g(t))$ of the front converges to zero uniformly in 
the following sense
$$
   \lim_{t \to +\infty} \sup_{x \in \real} \bigl( |f(x,t)| + |g(x,t)| 
   \bigr) \left(1+e^{sx}\right) \,=\, 0~.
$$
\noindent{\bf b)} In addition to \equ(limit), the proof will show that
$$
   \int_0^\infty \|(\partial_x f(t),\partial_x g(t))\|_{X_s^2}^2 \,dt 
   \,<\, \infty~, 
$$
but \clm(stab) does not give any pointwise decay rate in time for the
perturbations. However, our proof could be extended to provide some
(non-optimal) decay rate at the expense of using higher order energy 
functionals, see for example [FS]. On the other hand, a detailed study of 
the linearized equation \equ(fg) (with $N(f,g) = 0$) including a careful 
determination of the optimal decay rate in this case can be found in [Fo2]. 

\noindent{\bf c)} A important open problem is whether the result of 
\clm(stab) can be extended to other values of the parameters $D,k$, 
using possibly different perturbation spaces. 
The difficulty we encounter when $D$ is far from $1$ is related to
the ``Turing phenomenon'' in the theory of pattern formation [Tu], which 
shows how a stable equilibrium point of a reaction system can be destabilized 
by diffusion if the components in the system have very different 
diffusion rates. The question is therefore whether this mechanism actually 
leads to instabilities in our system if either $D \ll 1$ or $D \gg 1$. 
On the other hand, the difficulty we have when $k \gg 1$
is more technical in nature: the spectral analysis of the linearized 
operator in \equ(fg) becomes difficult when $k$ is large, because
we have to preclude the existence of unstable eigenvalues. This
problem already exists when $D=1$, but in this case the reduction to a single
equation allows one to use some results from the theory of Schr\"odinger
operators which do not extend to systems. Another interesting
question is therefore whether \clm(stab) holds true for all $k \ge 0$, at 
least if $D$ is close to $1$, and whether the orbital stability result for 
the family of translates of the ``pushed front'', which is known for $D=1$,
$k > 2$, $c = c^*(k)$, has any analogue in the general case $D \neq 1$. 

The rest of this paper is organized as follows. In section~2, we prove 
the existence of propagating fronts (\clm(Dlarge) and \clm(Dsmall))
using phase space techniques. In section~3, we construct energy functionals 
which allow us to show that these front solutions are stable against 
sufficiently small perturbations in $Y_s^2$ (\clm(stab)). 

\SECTION Existence of Propagating Fronts

In this section, we study the existence of nonnegative solutions to the 
system \equ(front) satisfying the boundary conditions \equ(bcond). 
Our method, which follows closely Billingham and Needham [BN], relies on the
construction of invariant regions in a three-dimensional phase space. 
When $D < 1$, $k \ge 0$, we show that a propagating front exists if $c$ 
is large enough, and does not exist if $c$ is too small, thus proving
\clm(Dsmall).  A similar result holds when $D > 1$, but in addition we can
show that the existence of a propagating front for $c > 0$, $k \ge 0$ 
implies the existence for all $c\pr \ge c$, $k\pr \le k$, thus proving 
\clm(Dlarge). 

We first note that any solution of \equ(front), \equ(bcond) corresponds to a 
heteroclinic orbit of a {\sl three-dimensional} dynamical system, see [BN]. 
Indeed, setting $\beta' = w$ and using the conservation law \equ(claw) to 
eliminate $\alpha'$, we see that $\alpha,\beta,w$ satisfy the equations
$$
  \alpha' \,=\, c(1-\alpha-\beta) - Dw~, \quad \beta' \,=\, w~, \quad  
  w' \,=\, -{1 \over D} (cw + \alpha\beta(1+k\beta))~, \EQ(3sys)
$$
together with the boundary conditions $(\alpha,\beta,w)(-\infty) = P_1 = 
(0,1,0)$, $(\alpha,\beta,w)(+\infty) = P_2 = (1,0,0)$. 

Straightforward calculations show that the fixed point $P_1$ is a {\sl saddle}
for the dynamical system \equ(3sys), with one positive and two negative 
eigenvalues given by
$$
   \lambda_{\pm} \,=\, -{c \over 2} \pm \sqrt{{c^2 \over 4} + 1 + k}~, \quad
   \lambda_0 \,=\, -{c \over D}~. \EQ(lam)
$$
The eigenvector corresponding to $\lambda_+ > 0$ is 
$$
   v_+ \,=\, \left({c+D\lambda_+ \over c+\lambda_+}~,~-1 ~,~ -\lambda_+
   \right)~. \EQ(eigvect)
$$
On the other hand, the fixed point $P_2$ is a {\sl sink}, with three negative
eigenvalues given by
$$
   \mu_{\pm} \,=\, {1 \over 2D}(-c \pm \sqrt{c^2-4D})~, \quad \mu_0 = -c~.
   \EQ(mupm)
$$
Note that $\mu_\pm > \mu_0$ if $D > 1/2$ and $1+(D-1)c^2 > 0$. The 
eigenvectors corresponding to $\mu_\pm$, $\mu_0$ are respectively
$$
   w_\pm \,=\, \left(-{c+D\mu_\pm \over c +\mu_\pm}~,~ 1 ~,~ \mu_\pm\right)~, 
   \quad w_0 \,=\, (1,0,0)~. \EQ(eigwect)
$$

Let $\VV$ be the one-dimensional global unstable manifold of $P_1$, and let
$\VV_+ \subset \VV$ be the invariant manifold which coincides with $\VV \cap
\QQ_+$ in a small neighborhood of $P_1$, where $\QQ_+ = \{(\alpha,\beta,w)\,|
\, \alpha > 0\,,\,\beta > 0\}$. In view of the preceding remarks, proving 
the existence of a nonnegative propagating front amounts to showing that 
$\VV_+ \subset \bar \QQ_+ \cap \Omega(P_2)$, where $\Omega(P_2)$ is the   
basin of attraction of $P_2$. Following [BN], we shall prove this, for some 
values of the parameters $D,c,k$, by constructing an invariant region 
$\dot \RR \subset \bar \QQ_+ \cap \Omega(P_2)$ such that $\VV_+ \subset \dot 
\RR$. Since everything is known when $D=1$, we shall only consider 
the cases $0 < D < 1$ and $D > 1$. 

\SUBSECTION The Case $D < 1$

Let $D \in (0,1)$. Given $A > 0$, we define the (closed) region $\RR_1 
\subset \bar \QQ_+$ by 
$$
   \RR_1 \,=\, \bigl\{ (\alpha,\beta,w) \,|\, 0 \le \beta \le 1~,~ 
   0 \le \alpha \le 1-\beta ~,~ -A\beta(1-\beta) \le w \le 0 \bigr\}~.
   \EQ(reg1)
$$
We also note $\dot \RR_1 = \RR_1 \setminus \{P_1\}$. 

\CLAIM Lemma(R1) Assume that the parameters $D < 1$, $c > 0$, 
$k \ge 0$, $A > 0$ satisfy
$$
   D A^2 - cA + 1 \,\le\, 0~, \quad k - 2D A^2 \,\le\, 0~. \EQ(inv1)
$$
Then $\RR_1$ is invariant under the flow of \equ(3sys), and $\VV_+ \subset
\dot \RR_1 \subset \Omega(P_2)$. 

\PROOF To prove the invariance of $\RR_1$, we show that the vector field
\equ(3sys) at any point $P \in \partial \RR_1$ is directed into $\RR_1$ or is 
parallel to the surface of $\RR_1$. This is easy to verify for the faces 
$\alpha = 0$, $\alpha = 1-\beta$, and $w = 0$. On the last face 
$w = -A\beta(1-\beta)$, we have
$$\eqalign{
  w' + A\beta'(1-2\beta) \,&=\, w\left({-c \over D}+A(1-2\beta)\right) 
    - {1 \over D}\alpha\beta(1+k\beta) \cr
  \,&=\, A\beta(1-\beta)\left({c \over D}- A(1-2\beta)\right) - {1 \over D}
    \alpha\beta(1+k\beta) \cr
  \,&\ge\, {1 \over D}\beta(1-\beta)\left(-(DA^2-cA+1)+\beta(2DA^2-k)\right)~,}
  \EQ(inf1)
$$
since $\alpha \le 1-\beta$. Obviously, the right-hand side of \equ(inf1) 
is nonnegative if \equ(inv1) holds, hence $\RR_1$ is invariant under the 
flow of \equ(3sys). 

Due to this invariance, to prove that $\VV_+ \subset \RR_1$ it suffices to 
verify the inclusion in a small neighborhood $\NN$ of $P_1$. Since $\VV_+ \cap
\NN \subset \{P_1 + \epsilon v_+ + \OO(\epsilon^2)\,|\, 0 < \epsilon < 
\epsilon_0\}$ for some $\epsilon_0 > 0$, it follows from \equ(eigvect), 
\equ(reg1) that $\VV_+ \cap \NN \subset \RR_1$ if
$$
   {c+D\lambda_+ \over c+\lambda_+} \,<\, 1~, \quad \hbox{and} \quad
   \lambda_+ \,<\, A~. \EQ(enter1)
$$
The first inequality is always satisfied since $D < 1$, $\lambda_+ > 0$, 
and the second one follows from \equ(lam), \equ(inv1). Indeed, adding the 
two inequalities \equ(inv1), we obtain $DA^2 + cA - (1+k) \ge 0$, while
$\lambda_+^2 + c\lambda_+ - (1+k) = 0$ by \equ(lam). Since $D < 1$, this
implies $A > \lambda_+$. Therefore, we have shown that $\VV_+ \subset \RR_1$, 
hence $\VV_+ \subset \dot \RR_1$ since $P_1 \notin \VV_+$.
   
Finally, let $P \in \dot \RR_1$, and let $\gamma(x) = (\alpha,\beta,w)(x)$ 
be the solution of \equ(3sys) satisfying $\gamma(0) = P$. Since
$\gamma(x) \in \RR_1$ for all $x \ge 0$, we have $\alpha'(x) \ge 0$, $\beta'(x)
\le 0$ for all $x \ge 0$, hence $\alpha(x) \to \tilde \alpha \in [0,1]$, 
$\beta(x) \to \tilde \beta \in [0,1)$ as $x \to +\infty$ (note that $\tilde 
\beta \le \beta(0) < 1$.) From \equ(3sys),
we deduce that $w(x) \to \tilde w = -c^{-1}\tilde\alpha\tilde\beta(1+k\tilde
\beta)$ as $x \to +\infty$. Now, the fixed point $(\tilde\alpha,\tilde\beta,
\tilde w)$ has to satisfy $\tilde w = 0$, $\tilde\alpha\tilde\beta = 0$ 
and $\tilde \alpha + \tilde \beta = 1$, hence $\tilde\alpha = 1$, 
$\tilde\beta = 0$. This proves that $\gamma(x) \to P_2$ as $x \to +\infty$,
hence $\dot \RR_1 \subset \Omega(P_2)$. \QED

\noindent{\bf Proof of \clm(Dsmall).}
{}From \clm(R1), we know that a nonnegative propagating front exists if the 
conditions \equ(inv1) can be satisfied for some $A > 0$. This front is 
unique up to a translation in $x$ since the unstable manifold $\VV_+$ is 
one-dimensional, and the properties \equ(prop) follow from \equ(reg1) or
can be verified directly as in [BN]. Now, if $k \le 2$, we choose $A = 
1/\sqrt{D}$, and \equ(inv1) holds for all $c \ge 2\sqrt{D}$. If $k > 2$, 
we choose $A = \sqrt{k/(2D)}$, and \equ(inv1) holds for all $c \ge \sqrt{D}
(\sqrt{k/2}+\sqrt{2/k})$. This proves the first part of the result. 

Conversely, if $c < 2\sqrt{D}$, the eigenvalues $\mu_\pm$ in \equ(mupm) 
become complex, and it is easy to show that no trajectory of \equ(3sys)
can stay in $\bar \QQ_+$ and converge to $P_2$ as $x \to +\infty$, 
except on the invariant line $\beta = w = 0$ which does not intersect
$\VV_+$. Therefore, no nonnegative front solution can exist if 
$c < 2\sqrt{D}$. This concludes the proof of \clm(Dsmall). \QED

\REMARK Following the same lines, one verifies that the region
$$
   \tilde \RR_1 \,=\, \bigl\{ (\alpha,\beta,w) \,|\, 0 \le \beta \le 1~,~ 
   (1-\beta)(1-E\beta) \le \alpha \le 1-\beta ~,~ -A\beta(1-\beta) \le w 
   \le 0 \bigr\}~
$$
satisfies the conclusion of \clm(R1) if $E > 0$ and if the additional 
condition $A(1-D)-(c-A)E \le 0$ is fulfilled (see also Lemma~2.2 below.) 
In particular, if $D > 1/2$, $c \ge 2\sqrt{D}$ and $k \le 2$, we can choose 
$A = 1/\sqrt{D}$ and $E = (1-D)/(2D-1)$. This shows that the propagating front
satisfies the bound
$$
   (1-\beta)\left(1 - {1-D \over 2D-1}\beta\right) \,\le\, \alpha \,\le\, 
   1-\beta~. \EQ(bdalpha1)
$$

\SUBSECTION The Case $D > 1$ : Bounds on the Critical Speed

Let $D >1$, $A > 0$, $E > 0$. We define the region $\RR_2 \subset \bar \QQ_+$
by 
$$
   \RR_2 \,=\, \bigl\{ (\alpha,\beta,w) \,|\, 0 \le \beta \le 1~,~ 
   1-\beta \le \alpha \le (1-\beta)(1+E\beta) ~,~ -A\beta(1-\beta) \le w 
   \le 0 \bigr\}~.
$$
We also note $\dot \RR_2 = \RR_2 \setminus \{P_1\}$. 

\CLAIM Lemma(R2) Assume that the parameters $D > 1$, $c > 0$, 
$k \ge 0$, $A > 0$, $E > 0$ satisfy
$$
   D A^2 - cA + 1 \,\le\, 0~, \quad E(1+k) + k - 2D A^2 \,\le\, 0~, \quad
   A(D-1) - E(c-A) \,\le\, 0~. \EQ(inv2)
$$
Then $\RR_2$ is invariant under the flow of \equ(3sys), and $\VV_+ \subset
\dot \RR_2 \subset \Omega(P_2)$. 

\PROOF We proceed as in the proof of \clm(R1). First, it is easy to verify
that the vector field \equ(3sys) is directed into $\RR_2$ on the faces 
$\alpha = 1-\beta$ and $w=0$. When $w=-A\beta(1-\beta)$, we have as in 
\equ(inf1)
$$\eqalign{
  w' + A\beta'(1-&2\beta) \,=\, A\beta(1-\beta)\left({c \over D}- A(1-2\beta)
   \right) - {1 \over D}\alpha\beta(1+k\beta) \cr
  \,&\ge\, {1 \over D}\beta(1-\beta)\left(cA-DA^2(1-2\beta)-(1+E\beta)
   (1+k\beta)\right) \cr
  \,&\ge\, {1 \over D}\beta(1-\beta)\left(-(DA^2-cA+1)+\beta(2DA^2-k-E(1+k))
   \right) \,\ge\, 0~,}
$$
since $\alpha \le (1-\beta)(1+E\beta)$ and $\beta^2 \le \beta$. Finally, when
$\alpha = (1-\beta)(1+E\beta)$, we find
$$\eqalign{
  \alpha' + \beta'(1-E+2E\beta) \,&=\, -cE\beta(1-\beta) + w(1-D-E+2E\beta)\cr
  \,&\le\, \beta(1-\beta)(-cE + A(D-1+E)) \,\le\, 0~,}
$$
since $-A\beta(1-\beta) \le w \le 0$. This proves that $\RR_2$ is invariant
under the flow of \equ(3sys). 

To show that $\VV_+ \subset \dot \RR_2$, it is sufficient to verify that 
$$
   1 \,<\, {c+D\lambda_+ \over c+\lambda_+} \,<\, 1+E~, \quad
   \hbox{and} \quad \lambda_+ < A~, \EQ(enter2), 
$$
see \equ(enter1). The first inequality is obvious since $D > 1$. To prove
that $\lambda_+ < A$, we multiply the last inequality in \equ(inv2) by $A$
and we add it to the sum of the other two; the result is 
$$
   A^2(1-E) + cA(1+E) - (1+k)(1+E) \,\ge\, 0~,
$$
hence $A^2 + cA - (1+k) > 0$, which implies $A > \lambda_+$. The second 
inequality in \equ(enter2) follows, since by \equ(inv2) $cE \ge A(D-1+E) 
 > \lambda_+(D-1-E)$. This proves that $\VV_+ \subset \dot \RR_2$. 

Finally, if $\gamma(x) = (\alpha,\beta,w)(x)$ is any trajectory of \equ(3sys)
in $\dot \RR_2$, then $\beta'(x) \le 0$ and $\alpha'(x) + D\beta'(x) \le 0$, 
hence $\beta(x)$ and $\alpha(x)$ converge as $x \to +\infty$. Thus, proceeding
as in the case $D < 1$, one shows that $\gamma(x) \to P_2$ as $x \to +\infty$. 
This proves that $\dot \RR_2 \subset \Omega(P_2)$. \QED

\clm(R2) ensures the existence of a nonnegative propagating front if the 
conditions \equ(inv2) can be satisfied for some $A > 0$, $E > 0$. These 
conditions are equivalent to
$$
   c \,\ge\, DA + 1/A \quad \hbox{and} \quad c \,\ge\, A\left(1+{D-1 \over 
   E}\right)~,
$$
for some $E \le (1+k)^{-1}(2DA^2-k)$. Therefore, given $D > 1$, $k \ge 0$, 
they can be fulfilled if and only if $c \ge \bar c(D,k)$, where
$$
   \bar c(D,k) \,=\, \min_{A^2 > k/(2D)} \max \left\{DA + 1/A~,~A\left(1+ 
   {(D-1)(1+k) \over 2DA^2 -k}\right)\right\}~. \EQ(cbar)
$$
In particular, setting $A^2 = 1/D$, we see that $\bar c(D,k) = 2\sqrt{D}$
if $k \le (3D-1)/(3D-2)$. Similarly, setting $A^2 = k/D$, we obtain 
$\bar c(D,k) \le \sqrt{D}(\sqrt{k}+1/\sqrt{k})$ for all $D > 1$, $k \ge 1$.
Finally, straightforward calculations show that
$$
   \lim_{D \to 1+} \bar c(D,k)  \,=\, c^*(k)~, \quad
   \lim_{D \to +\infty}{\bar c(D,k) \over \sqrt{D}} \,=\, c^*(2k)~,
$$
where $c^*(k)$ is given by \equ(cstar), and that $\DS\lim_{k \to +\infty} 
\bar c(D,k)/\sqrt{k}$ exists for all $D > 1$. This proves the upper bounds 
in \equ(bounds) for the critical speed $c^*(D,k)$. 

\REMARK If $k \le (3D-1)/(3D-2)$, the conditions \equ(inv2) are fulfilled 
for all $c \ge 2\sqrt{D}$ if $A = 1/\sqrt{D}$ and $E = (D-1)/(2D-1)$;
this shows that the propagating front satisfies the bound
$$
   (1-\beta) \,\le\, \alpha \,\le\, (1-\beta)\left(1+{D-1 \over 2D-1}\beta
   \right)~. \EQ(bdalpha2)
$$

To prove the lower bounds in \equ(bounds), we use a similar argument. First, 
we verify as in the case $D < 1$ that there exists no nonnegative propagating 
front if $c < 2\sqrt{D}$. Thus, we assume that $c \ge 2\sqrt{D}$, and we 
define the region $\RR_3 \subset \bar \QQ_+$ by
$$
   \RR_3 \,=\, \bigl\{ (\alpha,\beta,w) \,|\, 0 \le \beta \le 1~,~ 
   1-\beta \le \alpha~,~ w \le -B\beta(1-\beta)\bigr\}~, \EQ(reg3)
$$
for some $B > 0$. Then, if
$$
   DB^2 \,\ge\, 1~, \quad DB^2 -cB + 1 \,>\, 0~, \quad k -2DB^2 \,\ge\, 0~,
   \EQ(inv3)
$$
the vector field \equ(3sys) on $\partial\RR_3$ is always directed into 
$\RR_3$, except on the face $\beta=0$. Indeed, this is easy to verify for the
faces $\beta = 1$ and $\alpha = 1-\beta$. If $w = -B\beta(1-\beta)$, we have
as in \equ(inf1)
$$\eqalign{
  w' + B\beta'(1-2\beta) \,&=\, B\beta(1-\beta)\left({c \over D}-B(1-2\beta)
   \right) - {1 \over D}\alpha\beta(1+k\beta) \cr
  \,&\le\, {1 \over D}\beta(1-\beta)\left(-(DB^2-cB+1)+\beta(2DB^2-k)\right)
  \,\le\, 0~,}
$$
since $\alpha \ge 1-\beta$. In addition, the conditions \equ(inv3) ensure 
that the unstable manifold $\VV_+$ is contained in $\RR_3$ in a neighborhood
of $P_1$, namely
$$
   1 \,<\, {c+D\lambda_+ \over c+\lambda_+}~, \quad \hbox{and}\quad
   B \,<\, \lambda_+~,
$$
see \equ(enter1). The first inequality is obvious since $D > 1$, and the
second one follows by adding the last two inequalities in \equ(inv3)~: the
result is $DB^2 + cB - (1+k) \le 0$, hence $B^2 + cB - (1+k) < 0$, which
implies $B < \lambda_+$. 

According to these results, any trajectory on the unstable manifold $\VV_+$
either remains in $\RR_3$ for all $x \in \real$ or leaves $\RR_3$ (and
the positive sector $\bar \QQ_+$) by crossing the plane $\beta = 0$. Now, the
conditions \equ(inv3) also imply that no trajectory of \equ(3sys) can stay 
in $\RR_3$ and converge to $P_2$ as $x \to +\infty$, except on the invariant 
line $\beta = w = 0$ (which does not intersect $\VV_+$.) Indeed, since the 
eigenvalues \equ(mupm) satisfy $\mu_0 < \mu_\pm < 0$, any trajectory in
$\RR_3 \setminus \{\beta=w=0\}$ converging to $P_2$ becomes tangent to one
of the eigenvectors $w_\pm$ as $x \to +\infty$. In view of \equ(eigwect), 
\equ(reg3), this is possible only if $B \le |\mu_-|$, in contradiction with the
assumptions $DB^2 \ge 1$, $DB^2 -cB +1 > 0$ which imply $B >|\mu_-|$. 
Therefore, if \equ(inv3) holds, 
the invariant manifold $\VV_+$ necessarily crosses the plane $\beta = 0$ and 
no nonnegative propagating front can exist. In particular, if $k > 2$, we
set $B = \sqrt{k/(2D)}$, and we conclude from \equ(inv3) that no 
nonnegative propagating front exists if $c < \sqrt{D}(\sqrt{k/2} + 
\sqrt{2/k})$. This proves the lower bounds in \equ(bounds) for the critical 
speed $c^*(D,k)$.

\SUBSECTION The case $D > 1$ : Existence of the Critical Speed

Let $D > 1$. In this section, we show that the existence of a nonnegative 
propagating front for some value of the parameters $c,k$ implies the same
property for all $c\pr \ge c$, $k\pr \le k$. Thus we fix $c > 0$, $k \ge 0$, 
and we assume that $\alpha,\beta$ is a nonnegative solution of 
\equ(front), \equ(bcond). As in [BN], it is easy to verify that 
$\alpha'(x) > 0$, $\beta'(x) < 0$ and $\alpha(x) + \beta(x) > 1$ for all 
$x \in \real$. Setting $w(x) = \beta'(x)$ as usual, we consider the bounded 
region $\RR_4 \subset \bar \QQ_+$ delimited by the following four surfaces:

\item{} $S_1 = \{(\alpha,\beta,w)\,|\, w=0\}$~, \quad $S_2 = \{(\alpha,\beta,w)
\,|\, \alpha=1-\beta\}$~,

\item{} $S_3 = \{(\lambda\alpha(x),\beta(x),w(x))\,|\, x \in \real\,,~
0 \le \lambda \le 1\}$~,

\item{} $S_4 = \{(\alpha(x),\beta(x),\mu w(x))\,|\,x \in \real\,,~ 0 \le 
\mu \le 1\}$~. 

\noindent We also note $\dot \RR_4 = \RR_4 \setminus \{P_1\}$. 

\CLAIM Lemma(R4) For all $c\pr \ge c$, $0 \le k\pr \le k$, the region $\RR_4$ 
above is invariant under the flow of \equ(3sys)', and $\VV\prp \subset
\dot \RR_4 \subset \Omega\pr(P_2)$. 

\REMARK Here and in the sequel, \equ(3sys)' denotes the vector field 
\equ(3sys) with $c,k$ replaced by $c\pr,k\pr$, and similarly for $\VV\prp$ 
and $\Omega\pr(P_2)$. 

\PROOF We proceed as in the proof of \clm(R1) or \clm(R2). First, it is
straightforward to verify that the vector field \equ(3sys)' is directed
into $\RR_4$ on the surfaces $S_1$ and $S_2$. If $P \in S_3 \cap \partial
\RR_4$, then $P = (\lambda\alpha(x),\beta(x),w(x))$ for some $x \in \real$, 
$\lambda \in [0,1]$. Using the equations \equ(3sys) satisfied by $\alpha,
\beta,w$, it is easy to show that the vector
$$
   N_3(x,\lambda) \,=\, \bigl(0 ~,~ D^{-1}\alpha(cw+\alpha\beta(1+k\beta)) ~,~
   \alpha w\bigr)~,
$$
is normal to $S_3$ at $P$ and directed outside $\RR_4$. On the other hand, 
for any $c\pr,k\pr$ the vector field \equ(3sys)' at $P$ is given by
$$
   V\pra(x,\lambda) \,=\, \bigl(c\pr(1-\lambda\alpha-\beta)-Dw ~,~ w ~,~
   -D^{-1}(c\pr w + \lambda\alpha\beta(1+k\pr\beta))\bigr)~.
$$
Taking the scalar product, we thus obtain
$$
   N_3(x,\lambda) \cdot V\pra(x,\lambda) \,=\, {\alpha w \over D}
   \bigl((c-c\pr)w + (1-\lambda)\alpha\beta + (k-\lambda k\pr)\alpha \beta^2
   \bigr) \,\le\, 0~,
$$
since $\lambda \le 1$, $c\pr \ge c$ and $k\pr \le k$, hence the vector field 
\equ(3sys)' is directed into $\RR_4$ on $S_3 \cap \partial\RR_4$. Similarly, 
if $P = (\alpha(x),\beta(x),\mu w(x)) \in S_4$ for some $x \in \real$, 
$\mu \in [0,1]$, an exterior normal vector at $P$ is given by $N_4(x,\mu) =
(w^2 \,,\, Dw^2 - cw(1-\alpha-\beta) \,,\, 0)$, and the vector field 
\equ(3sys)' at $P$ reads
$$
   V\prb(x,\mu) \,=\, \bigl(c\pr(1-\alpha-\beta)-D\mu w ~,~ \mu w ~,~ 
   -D^{-1}(c\pr \mu w + \alpha\beta(1+k\pr\beta))\bigr)~.
$$
Therefore, we have $N_4(x,\mu) \cdot V\prb(x,\mu) = w^2(c\pr-c\mu)(1-\alpha
-\beta) \le 0$, since $\mu \le 1$ and $c\pr \ge c$. This shows that $\RR_4$ 
is invariant under the flow of \equ(3sys)' for all $c\pr \ge c$, $k\pr \le k$. 

To prove that $\VV\prp \subset \dot \RR_4$, we may clearly assume that
either $c\pr > c$ or $k\pr < k$, since $\VV_+ \subset \dot \RR_4$ by 
construction. Then, as in \equ(enter1), it is sufficient to verify that
$$
   1 \,<\, {c\pr + D\lambda\prp \over c\pr + \lambda\prp} \,<\, 
   {c + D\lambda_+ \over c+\lambda_+} ~, \quad \hbox{and }
   \lambda\prp \,<\, \lambda_+~,
$$
where $\lambda\prp = \lambda_+(c\pr,k\pr)$ is given by \equ(lam). The last
inequality is satisfied because $\lambda_+(c,k)$ is strictly decreasing in 
$c$ and increasing in $k$, and the other relations follow since $D > 1$. 
Therefore, $\VV\prp \subset \dot \RR_4$ for all $c\pr \ge c$, $k\pr \le k$. 

Finally, if $\bar \gamma(x) = (\bar \alpha,\bar \beta, \bar w)(x)$ is 
any trajectory of \equ(3sys)' in $\dot \RR_4$, then $\bar \beta'(x) \le 0$
and $\bar \alpha'(x) + D \bar \beta'(x) \le 0$, hence $\bar \beta(x)$ and
$\bar \alpha(x)$ converge as $x \to +\infty$. Proceeding as in the previous 
cases, one shows that $\bar \gamma(x) \to P_2$ as $x \to +\infty$. This proves
that $\dot \RR_4 \subset \Omega\pr(P_2)$. \QED

Using this result, we are now able to complete the proof of \clm(Dlarge). 

\noindent{\bf Proof of \clm(Dlarge).} For any $D > 1$, $k \ge 0$, let 
$I(D,k)$ be the set of values of $c \ge 2\sqrt{D}$ for which there exists
a nonnegative solution of \equ(front), \equ(bcond). It is not difficult to 
verify that this set is closed, hence by \clm(R4) $I(D,k) = [c^*,+\infty)$ 
for some $c^*(D,k) \ge 2\sqrt{D}$. 
It follows also from \clm(R4) that $I(D,k) \subset I(D,k\pr)$ if $k \ge k\pr$,
hence the minimal speed $c^*(D,k)$ is a nondecreasing function of $k \ge 0$. 
Finally, the upper and lower bounds \equ(bounds) for $c^*(D,k)$ have been 
established in Section~2.2. This concludes the proof of \clm(Dlarge). \QED

\SECTION Stability of the Propagating Fronts

Throughout this section, we assume that $D > 0$, $k \in [0,1]$, $c \ge 2
\sqrt{D}$, and we denote by $\alpha,\beta$ the solution of \equ(front), 
\equ(bcond) whose existence is ensured by \clm(Dlarge) or \clm(Dsmall). 
To prove \clm(stab), we shall control the behavior of the solutions of
\equ(fg) in the function space $Y_s^2$ defined by the norm \equ(ynorm),
\equ(schoice). We begin with a standard local existence result:

\CLAIM Lemma(local) Let $D > 0$, $k \in [0,1]$, $c \ge 2\sqrt{D}$,
and let $(f_0,g_0) \in Y_s^2$. Then there exists a time $t_1>0$ such that
\equ(fg) has a unique solution $(f,g) \in \CC^0([0,t_1], Y_s^2) \cap
\CC^1((0,t_1],Y_s^2)$ satisfying $(f(0),g(0)) = (f_0,g_0)$.

\PROOF It is straightforward to verify that the linear operator
$$
   \cal{A} \,=\, \left(\matrix{ \partial_x^2 + c \partial_x - \beta 
   (1 + k \beta) & - \alpha (1 + 2 k \beta)\cr \beta (1 + k \beta) &
   D \partial_x^2 + c \partial_x + \alpha (1 + 2k \beta)}\right)
$$
is the generator of an analytic semigroup in $Y_s^2$. Moreover, the 
nonlinearity $N : Y_s^2 \to Y_s$ in \equ(fg) is locally Lipschitz, uniformly
on any bounded subset of $Y_s^2$. Therefore, by Theorem~6.3.1 in [Pa], there 
exists a time $t_1 > 0$ such that Eq.\equ(fg) has a unique classical solution 
$(f,g) \in \CC^0([0,t_1], Y_s^2) \cap \CC^1((0,t_1],Y_s^2)$ with initial data 
$(f_0,g_0)$. \QED

\REMARK In addition, the proof shows that, for any bounded subset $B \subset
Y_s^2$, the existence time $t_1 > 0$ is bounded away from zero uniformly for 
all $(f_0,g_0) \in B$. It follows that the solution $(f,g)$ either exists for 
all $t \in \real_+$ or leaves any bounded subset of $Y_s^2$ in finite time. 

In the sequel, we fix $d_0 > 0$, $k_0 > 0$ sufficiently small, and 
we assume that $D \in [1-d_0,1+d_0]$, $k \in [0,k_0]$, $c \ge 2\sqrt{D}$. 
For $\epsilon > 0$ sufficiently small, we make the following assumption:

\noindent{\bf Hypothesis} $\HH_\epsilon$:
There exists a classical solution $(f,g)$ of \equ(fg) defined on some time 
interval $[0,T]$ and satisfying 
$$
  \|(f(t),g(t))\|_{Y_s^2} \,\le\, \epsilon~, 
$$
for all $t \in [0,T]$. 

Under this assumption, we shall study the time evolution of some energy 
functionals which control the size of the solution $(f,g)$ in $Y_s^2$.
In particular, if $\epsilon$, $d_0$, $k_0$ are sufficiently small, 
we shall show that the norm $\|(f,g)\|_{Y_s^2}$ remains bounded on any 
time interval $[0,T]$ by a quantity depending only on the initial data.
This result will be the main step in the proof of \clm(stab). 
As in [KR], [GR], it is convenient here to split the problem in two parts:
First, we shall construct {\sl weighted} functionals, with weight $e^{sx}$, 
which control the perturbations $(f,g)$ ahead of the propagating front. 
Then, we shall introduce {\sl unweighted} functionals to describe the 
behavior of $(f,g)$ behind the front. 

\SUBSECTION Weighted Functionals

Let $\rho(x) = e^{-sx}$, where $s$ is given by \equ(schoice). If $(f,g)$ is
any solution of \equ(fg) in $Y_s^2$, we define the weighted functions 
$F,G,H$ by
$$
   F(x,t) \,=\, \rho(x)^{-1} f(x,t)~, \quad
   G(x,t) \,=\, \rho(x)^{-1} g(x,t)~, \quad
   H \,=\, (1+ds^2)F + G~, \EQ(FGH)
$$
where $d = D-1$. Then $F,G,H \in \H^1(\real)$, and a direct calculation shows
that $G,H$ satisfy the system
$$\eqalign{
   \partial_t G \,=\, &(1+d)\partial_x^2 G + \mu \partial_x G 
      - (\hat{\gamma}+\hat{\beta})G + \hat{\beta}H
      + \rho\tilde{N}(G,H)~, \cr
   \partial_t H \,=\, &\partial_x^2 H + (\mu + 2ds)\partial_x H
      -(1 + ds^2)H + d(\partial_x^2 G-2s \partial_x G) \cr
      &+ ds^2 \left((\hat{\gamma}+\hat{\beta})G- \hat{\beta}H
      -\rho \tilde{N}(G,H)\right)~,} \EQ(GH)
$$
where 
$$ 
  \quad \mu \,=\, \sqrt{c^2-4D}~, \quad 
  \hat{\beta} \,=\, {\beta (1+k\beta) \over (1+ds^2)}~, \quad
  \hat{\gamma} \,=\, 1 - \alpha(1+2k\beta)~. \EQ(defchap)
$$ 
The nonlinearity $\tilde N$ in \equ(GH) is defined by
$$ 
  \tilde{N}(G,H) \,=\, \rho^{-2} N\left(\rho {H-G \over 1+ds^2}, \rho G
  \right)~. \EQ(tildeN)
$$

\REMARKS \HB
\noindent{\bf 1.}
Both functions $\hat \beta, \hat \gamma$ in \equ(defchap) are close to 
$\beta$ if $|d|$ and $k$ are sufficiently small. Indeed, since $s^2 \le 1/D 
= 1/(1+d)$ by \equ(schoice), we have $1+ds^2 = 1 + \OO(|d|)$, hence
$$
   \hat \beta/\beta \,=\, 1+\OO(|d|+k)~. \EQ(betachap)
$$
On the other hand, using the bounds \equ(bdalpha1), \equ(bdalpha2), it is 
straightforward to verify that
$$
   (1-|d|-2k)\beta \,\le\, \hat \gamma \,\le\, \beta\left(1+{|d| \over 1+2d}
   \right)~, \EQ(chap)
$$
if $d > -1/2$, hence $\hat \gamma/\beta = 1+\OO(|d|+k)$.
In the sequel, we shall always assume that $|d|$, $k$ are sufficiently small
so that $\hat \beta > 0$, $\hat \gamma > 0$ for all $x \in \real$.  

\noindent{\bf 2.} In the limit $x \to +\infty$, the equations \equ(GH) reduce
to
$$\eqalign{
   \partial_t G \,&=\, (1+d)\partial_x^2 G + \mu \partial_x G~, \cr
   \partial_t H \,&=\, \partial_x^2 H + (\mu + 2ds)\partial_x H
      -(1 + ds^2)H + d(\partial_x^2 G-2s \partial_x G)~.}
$$
For this limiting system, the ``energy'' $\int (G^2+H^2)dx$ is non-increasing 
in time if $d = D-1$ is sufficiently small. Indeed, the diagonal term 
$-(1+ds^2)H$ has a good sign, and the only cross term $d(\partial_x^2 
G-2s \partial_x G)$ is a derivative and is multiplied by the small parameter 
$d$. This almost diagonal form in the limit $x \to +\infty$ explains our 
choice of the variables $G,H$ instead of $F,G$. 

\noindent{\bf 3.} In the sequel, we shall use the estimate
$$
   {4 \over 3} \beta(x) \,\ge\, \cases{\rho(x) & if $x \ge 0$~,\cr
    1 & if $x \le 0$~,} \EQ(rhobeta)
$$
which holds for all $c \ge 2\sqrt{D}$ if $|d|$ and $k$ are sufficiently 
small. To prove \equ(rhobeta), we first note that $\beta'(x) + s\beta(x) > 0$
for all $x \in \real$. Indeed, using \equ(front), \equ(schoice), we see that
the function $z = \beta' + s\beta$ satisfies the equation
$$
   Dz' + (c-Ds) z + (\alpha + k\alpha\beta - 1)\beta \,=\, 0~.
$$
On the other hand, the bounds \equ(bdalpha1), \equ(bdalpha2) imply
that $\alpha +k\alpha\beta-1 \le \beta(-1+k+|d|)$, hence $Dz' + (c-Ds)z > 0$
for all $x \in \real$ if $k + |d| < 1$. Since $z(-\infty) = s > 0$, this 
differential inequality implies that $z(x) = \beta'(x) + s\beta(x) > 0$ 
for all $x \in \real$, hence $\beta(x)/\rho(x)$ is an increasing function. 
Now, recall that in the introduction we used the translation invariance 
of the problem to impose that $\beta(x)-\alpha(x) \ge 3/4$ for all $x \le 0$. 
In particular, we have $\beta(0) \ge 3/4$, hence $\beta(x)/\rho(x) \ge 
\beta(0) \ge 3/4$ if $x \ge 0$. Since $\beta$ is a decreasing function, 
we also have $\beta(x) \ge \beta(0) \ge 3/4$ for all $x \le 0$. This proves
\equ(rhobeta).  

To control the evolution of $G,H$ in $\H^1(\real)$, we introduce the energy 
functionals
$$ \eqalign{
   E_0(t) \,&=\, {1 \over 2}\int_{\real}(G^2 + H^2)\,dx~, \cr
   E_1(t) \,&=\, {1 \over 2}\int_{\real}\left(DG'^2 +(\hat{\gamma}+
    \hat{\beta})G^2 + H'^2 + (1+ds^2) H^2\right)\,dx~,}
$$
where $G' = \partial_x G$, $H' = \partial_x H$. 

\CLAIM Lemma(E0) There exist $d_0 > 0$, $k_0 > 0$ and $\epsilon > 0$ such that,
if the hypothesis $\HH_\epsilon$ above is satisfied, then there exists 
$K_1 > 0$ such that, for all $t \in (0, T]$,
$$
  \dot{E}_0(t) \,\le\, -K_1 E_1(t)~. \EQ(E0)
$$

\REMARKS \HB
\noindent{\bf 1.} The constant $K_1$ in \equ(E0) is independent of $d \in
[-d_0,d_0]$, $k \in [0,k_0]$, $c \ge 2\sqrt{D}$ and $T > 0$. 

\noindent{\bf 2.} Here and in the sequel, we denote by $\,\dot{}\,$ the time 
derivative to distinguish it from the space derivative $\,{}'\,$. Unless 
stated otherwise, all the integrals are taken over the whole real line $\real$.
 
\PROOF Since $(f,g) \in \CC^0([0,T],Y_s^2) \cap \CC^1((0,T],Y_s^2)$, we have 
$E_0 \in \CC^0([0,T]) \cap \CC^1((0,T])$. Using the first equation in \equ(GH) 
and integrating by parts, we obtain
$$\eqalign{
  {1 \over 2}{d \over dt}\int G^2\,dx \,&=\, 
   \int\left(-(1+d) G'^2 - (\hat{\gamma}+\hat{\beta}) G^2
   +\hat{\beta} GH +\rho\tilde{N}(G,H) G\right)\,dx \cr
  \,&\le\, \int\left(-(1+d) G'^2 - (\hat{\gamma}+\hat{\beta}/2) G^2
   +{1 \over 2}\hat{\beta}H^2 + \rho\tilde{N}(G,H) G\right)\,dx~.}
  \EQ(dtG)
$$
Similarly, using the second equation in \equ(GH) and integrating by parts, 
we have
$$\eqalign{
  {1 \over 2} {d \over dt}\int_{\real} H^2\,dx \,=\,& 
   \int \left(-H'^2 - (1+ds^2)H^2 - dG'H' - 2dsG'H\right)\,dx \cr
   \,&+\, ds^2 \int\left((\hat{\gamma}+\hat{\beta})GH - \hat{\beta} H^2 
   - \rho \tilde{N}(G,H)H\right)\,dx~.} 
$$
Since $-dG'H' \le |d|(G'^2+H'^2)/2$, $-2dsG'H \le |d|s(G'^2+H^2)$, 
$ds^2GH \le |d|s^2(G^2+H^2)/2$, and since $s^2 \le (1+d)^{-1}$, there 
exists $C_0 > 0$ such that
$$\eqalign{
  {1 \over 2} {d \over dt}\int_{\real} H^2\,dx \,\le\,& 
   -\int (H'^2+H^2)\,dx - ds^2 \int \rho \tilde{N}(G,H)H\,dx \cr
  \,&+\, C_0 |d| \int\left(H'^2+H^2+G'^2+(\hat\gamma + \hat\beta)(G^2+H^2)
   \right)\,dx~.} \EQ(dtH)
$$

To bound the nonlinear terms in \equ(dtG), \equ(dtH), we first note that, 
due to the hypothesis $\HH_\epsilon$, there exists $C_1>0$ such that
$$
   \rho(x) |G(x,t)| \,\le\, C_1 \epsilon \beta(x)~, \EQ(rhoG)
$$ 
for all $x \in \real$, $t \in [0,T]$. Indeed, if $x \le 0$, using \equ(FGH), 
\equ(rhobeta) and the embedding of $\H^1(\real)$ into $\L^\infty(\real)$, we 
find
$$
   \rho(x)^2 |G(x)|^2 \,=\, |g(x)|^2 \,\le\, \|g\|_{\infty}^2 
   \,\le\, \|g\|_2 \|g'\|_2 \,\le\, {1 \over 2} \|g\|_{Y_s}^2 
   \,\le\, {8 \over 9}\beta(x)^2 \|g\|_{Y_s}^2~.
$$
If $x \ge 0$, we have by \equ(rhobeta)
$$
   \rho(x)^2 |G(x)|^2 \,\le\, {16 \over 9} \beta(x)^2 |G(x)|^2
   \,\le\, {16 \over 9} \beta(x)^2 \|G\|_{\infty}^2 \,\le\, 
   {16 \over 9} \beta(x)^2 \|G\|_2 \|G'\|_2~.
$$
Since $|G'(x)| \le (s|g(x)|+|g'(x)|)e^{sx}$, we have $\|G\|_2 \|G'\|_2 \le 
C \|g\|_{Y_s}^2$ for some $C > 0$. Therefore, there exists $C_1 > 0$ such that
$\rho(x)|G(x)| \le C_1 \beta(x) \|g\|_{Y_s}$ for all $x \in \real$, and 
\equ(rhoG) follows from the hypothesis $\HH_\epsilon$. 

Now, using \equ(tildeN), \equ(fg), we have
$$
   \int \rho|\tilde{N}(G,H) G|\,dx \,\le\, 
   \int \rho |G| \left({|H-G||G| \over 1+ds^2}
   (1 + 2k\beta +k\rho |G|) + k \alpha G^2 \right)\,dx~.
$$
Since $|H-G||G| \le (3G^2+H^2)/2$ and $(1+ds^2)^{-1} = 1+\OO(|d|)$, it follows
from \equ(rhoG) that there exists $C_2 > 0$ such that
$$\eqalign{
  \int \rho|\tilde{N}(G,H) G|\,dx 
  \,&\le\, C_1 \epsilon \int \beta \left({3 G^2 + H^2 \over 2(1+ds^2)}
    (1+2k+kC_1\epsilon)+kG^2\right)\,dx \cr
  \,&\le\, C_2 \epsilon \int \beta(G^2+H^2)\,dx~.} \EQ(N1a)
$$
In a similar way, we obtain
$$
  \int \rho|\tilde{N}(G,H) H|\,dx\,\le\,  
  C_2 \epsilon \int \beta(G^2 + H^2)\,dx~. \EQ(N1b)
$$
Therefore, combining \equ(dtG), \equ(dtH), \equ(N1a), \equ(N1b), and using
\equ(betachap), \equ(chap), we see that there exists $C_3 > 0$ such that
$$\eqalign{
  \dot E_0 \,\le\, &-\int \left(G'^2 + (\hat\gamma + \hat\beta/2)G^2 + H'^2 + 
   {1 \over 2}H^2\right)\,dx \cr
  &+ C_3(|d|+k+\epsilon) \int \left(G'^2 + (\hat\gamma + \hat\beta)G^2
   + H'^2 + H^2 \right)\,dx~.}
$$
In particular, if $|d|$, $k$ and $\epsilon$ are sufficiently small, there 
exists $K_1>0$ (independent of $c \ge 2\sqrt{D}$ and $T>0$) such that
$$
   \dot E_0 \,\le\, - {K_1 \over 2} \int \left(DG'^2
    +(\hat{\gamma}+\hat{\beta})G^2 + H'^2 + (1+ds^2) H^2\right)\,dx
   \,=\, - K_1 E_1~.
$$
This concludes the proof of \clm(E0). \QED

\CLAIM Lemma(E1) There exist $d_0 > 0$, $k_0 > 0$ and $\epsilon > 0$ such 
that, if the hypothesis $\HH_\epsilon$ is satisfied, then
there exists $K_2(c) > 0$ such that, for all $t \in (0, T]$,
$$
  \dot{E}_1(t) \,\le\, K_2(c) E_1(t)~.\EQ(E1)
$$

\REMARK The constant $K_2(c)$ is independent of $T>0$, and behaves like $c^2$ 
for large values of $c$, uniformly in $d \in [-d_0, d_0]$ and $k \in [0, k_0]$.

\PROOF We start from the identity
$$
  I_1 \,\equiv\, {1 \over 2} {d \over dt} \int\left((1+d)G'^2
  +(\hat{\gamma}+\hat{\beta})G^2\right)\,dx
  \,=\, \int \left(-(1+d) G''+(\hat{\gamma}+\hat{\beta})G\right)\dot{G}\,dx~.
$$
Using \equ(GH), we replace $-(1+d) G''+(\hat{\gamma}+\hat{\beta})G$ with 
$-\dot{G}+\mu G'+ \hat{\beta}H + \rho \tilde{N}(G,H)$ in the right-hand side. 
We obtain
$$
  I_1 \,=\, \int \left(-\dot{G}^2 + \mu G'\dot{G} + \hat{\beta}H\dot{G}
  + \rho \tilde{N}(G,H) \dot{G}\right)\,dx~.
$$
Since $\mu G'\dot{G} \le \mu^2 G'^2 + \dot{G}^2/4$ and $\hat{\beta} H\dot{G}
\le \hat{\beta}(H^2+\dot{G}^2)/2$, we have
$$
  I_1 \,\le\, \int \left(-{3 \over 4}\dot{G}^2 + \mu^2 G'^2 + {1 \over 2}
  \hat{\beta}(H^2 + \dot{G}^2) + \rho\tilde{N}(G,H) \dot{G}\right)\,dx~.
  \EQ(dtGprime)
$$

Similarly, we have the identity
$$ 
  I_2 \,\equiv\, {1 \over 2} {d \over dt} \int \left(H'^2 + (1+ds^2)H^2\right)
  \,dx \,=\, \int \left(-H'' + (1+ds^2)H\right)\dot{H}\,dx~.
$$
Replacing $-H'' + (1+ds^2)H$ in the right-hand side with its expression 
obtained from the second equation in \equ(GH), we find
$$ \eqalign{
  I_2 \,=\, &\int \left(-{\dot H}^2 + (\mu+2ds)H'{\dot H} + dG''{\dot H}
   -2ds G' {\dot H}\right)\,dx \cr
  &+ ds^2 \int\left((\hat\gamma + \hat\beta)G - \hat\beta H - \rho\tilde{N}
   (G,H)\right){\dot H}\,dx~.}
$$
In view of \equ(GH), we also have
$$
   G'' \,=\, {1 \over (1+d)} \left(\dot{G} - \mu G' + (\hat{\gamma}+\hat
   {\beta})G -\hat{\beta}H-\rho\tilde{N}(G,H)\right)~. \EQ(Gprpr)
$$
Replacing \equ(Gprpr) into the expression of $I_2$, we find
$$\eqalign{
  I_2 \,=\, &\int \left(-{\dot H}^2 + (\mu+2ds) H'\dot{H} - 2ds G'\dot{H}
   + ds^2 (\hat{\gamma}+\hat{\beta})G{\dot H} - ds^2\hat{\beta}H\dot{H}\right)
   \,dx \cr
  &+ {d \over 1+d} \int\left(\dot{G} -\mu G' +(\hat{\gamma}+\hat{\beta})G 
   -\hat{\beta}H\right)\dot{H}\,dx \cr
  &-\left({d \over 1+d}+ds^2 \right)\int\rho\tilde{N}(G,H)\dot{H}\,dx~.}
  \EQ(dtHprime1)
$$
Since $(\mu+2ds)H'{\dot H} \le {\dot H}^2/2 + (\mu^2 + 4d^2s^2)H'^2$ and
$|\mu G'\dot H| \le \mu^2 G'^2/2 + {\dot H}^2/2$, it is easy to verify that
there exists $C_4 > 0$ such that
$$\eqalign{
  I_2 \,\le\, &\int\left(-{1 \over 2}{\dot H}^2 + \mu^2 H'^2\right)\,dx + 
   C_4|d| \int \rho|\tilde{N}(G,H)\dot H|\,dx \cr
  &+ C_4 |d| \int\left({\dot H}^2 + H'^2 + H^2 + {\dot G}^2 + 
   (1+\mu^2)G'^2 + (\hat\gamma + \hat \beta)G^2\right)\,dx~.} \EQ(dtHprime)
$$

To bound the nonlinear terms in \equ(dtGprime), \equ(dtHprime), we proceed
as in the proof of \clm(E0). We obtain
$$\eqalign{
  \int \rho|\tilde{N}(G,H) {\dot G}|\,dx \,&\le\,  
    C_2 \epsilon \int \beta(G^2 + {\dot G}^2 + H^2)\,dx~, \cr 
  \int \rho|\tilde{N}(G,H) {\dot H}|\,dx \,&\le\,  
    C_2 \epsilon \int \beta(G^2 + {\dot H}^2 + H^2)\,dx~.} \EQ(N2) 
$$
Therefore, combining \equ(dtGprime), \equ(dtHprime), \equ(N2) and 
using \equ(betachap), \equ(chap), we see that there exists $C_5 > 0$ 
such that
$$\eqalign{
  \dot E_1 \,\le\, &\int\left( -{1 \over 4}{\dot G}^2 -{1 \over 2}{\dot H}^2
   +\mu^2 (G'^2+H'^2) +{1 \over 2}H^2 \right)\,dx \cr
  &+ C_5(|d|+k+\epsilon) \int\left({\dot H}^2 + H'^2 + H^2 + {\dot G}^2
   + (1+\mu^2)G'^2 + (\hat\gamma + \hat\beta)G^2\right)\,dx~.}
$$
In particular, if $|d|$, $k$ and $\epsilon$ are sufficiently small, there 
exists $C_6 > 0$ such that $\dot E_1 \,\le\, C_6(1+\mu^2)E_1$, hence \equ(E0) 
holds with $K_2(c) = C_6(1+\mu^2)$. This concludes the proof of \clm(E1). \QED

\SUBSECTION Unweighted Functionals

To control the perturbation $(f,g)$ behind the front, we define $h = f + g$,
and we consider the equations satisfied by $f,h$. From \equ(fg), we obtain
$$\eqalign{
  \partial_t f \,=\, \phantom{D}& \partial_x^2 f + c\partial_x f - \delta f 
               -\alpha(1+2k\beta)h - N(f, h-f)~, \cr
  \partial_t h \,=\, D& \partial_x^2 h + c\partial_x h - d\partial_x^2 f~,}
  \EQ(fh)
$$
where $\delta = \beta-\alpha + k\beta(\beta - 2\alpha)$. 
As in the weighted case, the variables $f,h$ have been chosen so that the
system \equ(fh) becomes almost diagonal in the limit $x \to -\infty$. To
control the evolution of $f,h$ in $\H^1(\real)$, we define the functionals
$$ \eqalign{
  E_2(t) \,&=\, {1 \over 2}\int_{\real}(f^2 + h^2 + 2\alpha^2 h^2)\,dx~, \cr
  E_3(t) \,&=\, {1 \over 2}\int_{\real}(f'^2+h'^2) 
            + {K \over 2}\int_{\real}(f^2 + 2\alpha^2 h^2)\,dx~,}
$$
where $K > 6$ is an arbitrary constant.   

Before computing the time derivative of $E_2, E_3$, we note that the 
additional term $\int \alpha^2 h^2 dx$ in $E_2$ satisfies
$$
   {d \over dt} \int \alpha^2 h^2 \,dx \,=\, -2D \int \alpha^2 {h'}^2\,dx
   + 2 \int \sigma h^2\,dx - 2d \int \alpha^2 f''h \,dx~, \EQ(alphah)
$$
where $\sigma(x) = (D(\alpha^2)''-c(\alpha^2)')/2$. In the sequel, we shall
use the following two properties of $\sigma$:

\noindent{i)} Since $\alpha(x) \sim e^{\lambda_+ x}$ as $x \to -\infty$, 
where $\lambda_+$ is given by \equ(lam), we have
$$\eqalign{
   \lim_{x \to -\infty} {\sigma(x) \over \alpha^2(x)} \,&=\, 2D\lambda_+^2 
    -c\lambda_+ \,=\, (2D+1)\lambda_+^2 -1-k \cr
   \,&\le\, (2D+1)(\sqrt{D+1+k}-\sqrt{D})^2 -1-k \cr
   \,&=\, -2\sqrt{2}(\sqrt{2}-1)^2 + \OO(|d|+k)~,}
$$
where we used the fact that $\lambda_+ \le \sqrt{D+1+k}-\sqrt{D}$ for all
$c \ge 2\sqrt{D}$. Therefore, if $|d|$ and $k$ are sufficiently small,
there exists $x_0 \in \real$ such that
$\sigma(x) \le -\alpha^2(x)/3$ for all $x \le x_0$.
Using the translation invariance of the problem, we may assume (without
loss of generality) that $x_0 \ge 0$, hence
$$
   \sigma(x) \,\le\, -{\alpha^2(x) \over 3}~, \quad \hbox{for all } x 
   \,\le\, 0~. \EQ(sigma)
$$
This condition is obviously compatible with the previous requirement that 
$\beta(x) - \alpha(x) \ge 3/4$ when $x \le 0$. 

\noindent{ii)} For all $x \in \real$, we have
$$
   \sigma(x) \,\le\, 2D(1+k)\alpha^2(x)~. \EQ(sigma2)
$$
Indeed, let $z(x) = \alpha'(x)-\sqrt{1+k}\,\alpha(x)$. Then $z(x) \sim 
e^{\lambda_+ x}(\lambda_+ - \sqrt{1+k})$ as $x \to -\infty$. Since $\lambda_+
< \sqrt{1+k}$ by \equ(lam), it follows that $z(x) < 0$ if $x$ is sufficiently 
negative. On the other hand, in view of \equ(front), the function $z$ 
satisfies the differential inequality
$$\eqalign{
   z' + (c+\sqrt{1+k})z \,&=\, \alpha'' + c\alpha' -(1+k)\alpha -c\sqrt{1+k}
    \,\alpha \cr
   \,&=\, -\alpha(1-\beta)(1+k+k\beta) -c\sqrt{1+k}\,\alpha \,<\, 0~,}
$$
which implies that $z(x)$ stays negative for all $x \in \real$, hence 
$0 < \alpha'(x) < \sqrt{1+k}\,\alpha(x)$ for all $x \in \real$. Using
\equ(front) again, we conclude
$$ 
  \sigma \,=\, D\alpha'^2 + D\alpha\alpha'' - c\alpha\alpha'
  \,=\, D\alpha'^2 + D\alpha^2\beta(1+k\beta) - c(1+D)\alpha\alpha'
  \,\le\, 2D(1+k)\alpha^2~,
$$
which proves \equ(sigma2). 

Now, we shall control the time evolution of $E_2$ and $E_3$. 

\CLAIM Lemma(E2) There exist $d_0 > 0$, $k_0 > 0$ and $\epsilon > 0$ such 
that, if the hypothesis $\HH_\epsilon$ is satisfied, then
there exist $K_3 > 0$ and $K_4 > 0$ such that, for all $t \in (0, T]$,
$$
  \dot{E}_2(t) \,\le\, -K_3 E_3(t) + K_4 E_1(t)~. \EQ(E2)
$$

\REMARK The constants $K_3$ and $K_4$ are independent of $d \in
[-d_0,d_0]$, $k \in [0,k_0]$, $c \ge 2\sqrt{D}$ and $T > 0$. 

\PROOF Using \equ(fh) and integrating by parts, we obtain
$$\eqalign{
  {1 \over 2} {d \over dt}\int f^2\,dx \,&=\, \int\left(-f'^2 -\delta f^2
   -\alpha(1+2k\beta)fh - N(f,h-f)f\right)\,dx \cr
  \,&\le\, \int\left(-f'^2 -\delta f^2 + ({1 \over 2}+k\beta)
   (f^2 + \alpha^2h^2) - N(f,h-f)f\right)\,dx~.} \EQ(dtf1)
$$
Since $-\delta = \alpha-\beta + k\beta(2\alpha-\beta) \le \alpha-\beta +
k\alpha^2 \le \alpha-\beta + k$, we have $-\delta \le 1+k$ for all $x \in 
\real$ and, due to our choice of the origin, $-\delta \le -3/4 + k$ for all 
$x \le 0$. Therefore, 
$$\eqalign{
  \int\left({1 \over 2}+k\beta-\delta\right)f^2\,dx \,&\le\, 
   \int_{-\infty}^0 \left(-{1 \over 4}+2k\right)f^2\,dx + \int_0^{\infty}
   \left({3 \over 2} + 2k\right)f^2\,dx \cr
  \,&=\, \int_\real\left(-{1 \over 4}+2k\right)f^2\,dx + {7 \over 4} 
   \int_0^{\infty} f^2\,dx~.} \EQ(betaf) 
$$
To bound the nonlinear term in \equ(dtf1), we observe that
$$\eqalign{
  |N(f,h-f)f| \,&=\, \left| f^2(h-f)(1+2k\beta) + k\alpha f(h-f)^2 + 
   kf^2(h-f)^2 \right| \cr
  \,&\le\, \|h-f\|_\infty \left(f^2(1+2k+k\|h-f\|_\infty)+k\alpha|f(h-f)|
   \right)~.}
$$
Since $\alpha|f(h-f)| \le \alpha f^2 + (f^2+\alpha^2 h^2)/2$ and since 
$\|h-f\|_\infty = \|g\|_\infty \le \|g\|_{Y_s} \le \epsilon$ by the hypothesis
$\HH_\epsilon$, there exists $C_1 > 0$ such that
$$
   \int |N(f, h-f)f|\,dx \,\le\, C_1 \epsilon \int(f^2 + 
   \alpha^2 h^2)\,dx~. \EQ(N3)
$$
Combining \equ(dtf1), \equ(betaf), \equ(N3), we obtain
$$\eqalign{
  {1 \over 2} {d \over dt}\int f^2\,dx \,\le\, &\int \left(-f'^2 -{1 \over 4}
   f^2 + {1 \over 2}\alpha^2 h^2\right)\,dx + (C_1 \epsilon + 2k) \int (f^2
   +\alpha^2 h^2)\,dx \cr
  &+ {7 \over 4} \int_0^{\infty} f^2\,dx~.} \EQ(dtf)
$$

Next, using \equ(fh) and integrating by parts, we find
$$
  {1 \over 2} {d \over dt}\int h^2\,dx \,=\, \int (-Dh'^2 + df'h')\,dx
  \,\le\, \int\left( \bigl(-D+{|d| \over 2}\bigr)h'^2 + {|d| \over 2}f'^2
  \right)\,dx~. \EQ(dth)
$$
Finally, integrating by parts in \equ(alphah), we obtain
$$\eqalign{
  {d \over dt} \int \alpha^2 h^2 \,dx \,&=\, \int \left(-2D \alpha^2 h'^2
   + 2 \sigma h^2 + 2d \alpha^2 f'h' + 2d(\alpha^2)'f'h \right)\,dx\cr
  \,&\le\, \int \left((-2D+|d|)\alpha^2 h'^2 + (2\sigma+|d|(\alpha^2)')h^2 
   + |d| (\alpha^2+(\alpha^2)')f'^2\right)\,dx~.} \EQ(dtalphah)
$$
In view of \equ(sigma), \equ(sigma2), we have
$$\eqalign{
  2\int_{\real}\sigma h^2\,dx \,&\le\, -{2 \over 3}\int_{-\infty}^0
   \alpha^2 h^2\,dx + 4D(1+k)\int_0^{\infty}\alpha^2 h^2\,dx \cr
  \,&\le\, -\,{2 \over 3}\int_{\real} \alpha^2h^2\,dx + \left({2 \over 3}
   +4D(1+k)\right)\int_0^{\infty}\alpha^2h^2\,dx~.} \EQ(sigmah)
$$
Replacing \equ(sigmah) into \equ(dtalphah) and recalling that $(\alpha^2)' 
= 2\alpha\alpha' \le 2\sqrt{1+k}\,\alpha^2$, we thus find
$$\eqalign{
  {d \over dt} \int \alpha^2 h^2 \,dx \,\le\, &-\int\left(2\alpha^2 h'^2
   +{2 \over 3}\alpha^2 h^2\right)\,dx + 5 \int_0^\infty \alpha^2 h^2 \,dx \cr
  &+ C_2(|d|+k) \int\alpha^2(h'^2 + h^2 +f'^2)\,dx~,} \EQ(dtalphah2)
$$
for some $C_2 > 0$. 

Therefore, combining \equ(dtf), \equ(dth), \equ(dtalphah2), we see that there
exists $C_3 > 0$ such that
$$\eqalign{
  \dot{E}_2 \,&\le\, -\int \left(f'^2 + {1 \over 4}f^2 + {1 \over 6}\alpha^2
   h^2 + (1+2\alpha^2)h'^2 \right)\,dx \cr
  \,&+\, C_3(|d|+k+\epsilon)\int(f'^2 + f^2 + h'^2 + \alpha^2 h^2)\,dx
   + 5 \int_0^\infty (f^2 + \alpha^2 h^2)\,dx~.} \EQ(dtE2)
$$
In particular, if $|d|$, $k$ and $\epsilon$ are sufficiently small, there 
exists $K_3 > 0$ (depending on $K$) such that
$$
  \dot{E}_2 \,\le\, -K_3 E_3 + 5 \int_0^{\infty} (f^2 + h^2)\,dx~.
$$
It remains to show that $\int_0^\infty (f^2+h^2)\,dx \le C E_1$ for some
$C > 0$. Using \equ(FGH), \equ(rhobeta), we have for all $x \ge 0$
$$
   f^2 + h^2 \,\le\, 3 (f^2 + g^2) \,=\, 3 \rho^2 (F^2 + G^2)
   \,\le\, 3\rho (F^2+G^2) \,\le\, 4\beta (F^2+G^2)~.
$$
Thus, if $|d|$ and $k$ are sufficiently small, it follows from \equ(FGH), 
\equ(betachap) that there exist $C_4 > 0$, $C_5 > 0$ such that $f^2 + h^2 
\le C_4 \beta(G^2 + H^2) \le C_5(\hat \beta G^2 + (1+ds^2)H^2)$, 
for all $x \ge 0$, hence $\int_0^\infty (f^2+h^2)\,dx \le 2C_5 E_1$. 
This concludes the proof of \clm(E2). \QED

\CLAIM Lemma(E3) There exist $d_0 > 0$, $k_0 > 0$ and $\epsilon > 0$ such 
that, if the hypothesis $\HH_\epsilon$ is satisfied, then for all 
$t \in (0, T]$,
$$
  \dot{E}_3(t) \,\le\, K K_4 E_1(t)~. \EQ(E3)
$$

\PROOF Using \equ(fh) and integrating by parts, we obtain
$$
  {1 \over 2}{d \over dt}\int f'^2\,dx \,=\, \int\left(-f''^2 + \delta ff''
  +\alpha(1+2k\beta)h f'' + N(f,h-f)f''\right)\,dx~.
$$
Since $|\delta| \le 1+k$, we have $\delta ff'' \le f''^2/4 + (1+k)^2f^2$ and
$\alpha(1+2k\beta)hf'' \le f''^2/4 + \alpha^2(1+2k)^2h^2$. Moreover, as in 
\equ(N3), we have
$$
   \int |N(f, h-f)f''|\,dx \,\le\, C_1 \epsilon \int(f''^2 + f^2 + 
   \alpha^2 h^2)\,dx~. 
$$
Therefore, there exists $C_6 > 0$ such that
$$
  {1 \over 2}{d \over dt}\int f'^2\,dx \,\le\, \int\left(-{1 \over 2}f''^2
  + f^2 + \alpha^2 h^2\right)\,dx + C_6(\epsilon+k)\int(f''^2 + f^2 + 
  \alpha^2 h^2)\,dx~. \EQ(dtfprime)
$$
Similarly, using \equ(fh) and integrating by parts, we find
$$
  {1 \over 2}{d \over dt}\int h'^2\,dx \,=\, \int(-Dh''^2 + df''h'')\,dx
  \,\le\, -D \int h''^2\,dx + {|d| \over 2} \int(f''^2+h''^2)\,dx~.
$$
Combining these results, we see that, if $|d|$, $k$ and $\epsilon$ are 
sufficiently small, then
$$
  {1 \over 2}{d \over dt}\int (f'^2 + h'^2)\,dx \,\le\, 
  (1+C_6(\epsilon+k)) \int(f^2+\alpha^2 h^2)\,dx~. \EQ(dtfhprime)
$$

On the other hand, using \equ(dtf) and \equ(dtalphah2), we obtain as in
\equ(dtE2)
$$\eqalign{
  {K \over 2} &{d \over dt}\int(f^2+2\alpha^2 h^2)\,dx 
   \,\le\, -K\int \left(f'^2 + {1 \over 4}f^2 + {1 \over 6}\alpha^2
   h^2 + 2\alpha^2 h'^2 \right)\,dx \cr
  \,&+\, C_7(|d|+k+\epsilon)\int(f'^2 + f^2 + \alpha^2 h'^2 + \alpha^2 h^2)\,dx
   + 5K \int_0^\infty (f^2 + \alpha^2 h^2)\,dx~,} 
$$
for some $C_7 > 0$. Since $K > 6$ by assumption, it follows that, if $|d|$, 
$k$ and $\epsilon$ are sufficiently small, then
$$
   \dot E_3 \,=\, {1 \over 2}{d \over dt} \int (f'^2 + h'^2)\,dx + {K \over 2}
   {d \over dt}\int(f^2+\alpha^2 h^2)\,dx \,\le\, 5K \int_0^\infty
   (f^2+\alpha^2 h^2)\,dx~.
$$
Proceeding as in the proof of \clm(E3), we thus obtain $\dot E_3 \le KK_4
E_1$. This concludes the proof of \clm(E3). \QED

\SUBSECTION Proof of \clm(stab).

We first state two Corollaries which are direct consequences of the
preceding Lemmas.

\CLAIM Corollary(B) There exist $d_0 > 0$, $k_0 > 0$ and $\epsilon > 0$ 
such that, if the hypothesis $\HH_\epsilon$ is satisfied, then there exists 
$K_0(c) > 0$ independent of $T$ such that, for all $t \in [0, T]$,
$$
  \|(f(t),g(t))\|_{Y_s^2} \le K_0(c) \|(f_0,g_0)\|_{Y_s^2}~. \EQ(gbound)
$$

\PROOF According to the four preceding Lemmas, we can choose $d_0 > 0$, 
$k_0 > 0$ and $\epsilon > 0$ sufficiently small so that, if the hypothesis 
$\HH_\epsilon$ is satisfied, then the differential inequalities \equ(E0), 
\equ(E1), \equ(E2) and \equ(E3) hold for $t\in [0,T]$. In particular, the 
function 
$E_4\equiv K_1(E_1 + E_2 + E_3) + (K_2 + K_4 + K K_4)E_0$ is non-increasing 
in time for $t\in [0,T]$. On the other hand, it is straightforward to verify 
that there exist $K_5 > 0$ and $K_6 > 0$ (independent of $c$, $T$) such that
$$
   K_5 \|(f,g)\|_{Y_s^2}^2 \,\le\, E_4 \,\le\, K_6 \|(f,g)\|_{Y_s^2}^2~,
$$
hence \equ(gbound) holds with $K_0 = (K_6/K_5)^{1/2}$. Since $K_2 = \OO(c^2)$ 
as $c \to +\infty$, the same is true for $K_6$, hence $K_0 = \OO(c)$ as
$c \to +\infty$. \QED

\CLAIM Corollary(H) There exist $d_0 > 0$, $k_0 > 0$ and $\epsilon > 0$ such
that, if the hypothesis $\HH_\epsilon$ holds for all $T > 0$, then the solution
$(f,g)$ of \equ(fg) satisfies
$$
   \lim_{t \to +\infty}\|(\partial_x f(t),\partial_x g(t))\|_{X_s^2} \,=\, 0~.
   \EQ(gdecay)
$$

\PROOF We choose $d_0 > 0$, $k_0 > 0$ and $\epsilon > 0$ as in \clm(B). 
Using \equ(E0) and \equ(E1), we first note that the positive functions
$E_0(t)$ and $K_1 E_1(t) + K_2 E_0(t)$ are non-increasing in time, hence
converge as $t \to +\infty$. In particular, $E_1(t)$ has a nonnegative limit
as $t \to +\infty$. Now, by \equ(E0), we have
$$
   \int_0^{+\infty}E_1(t)\,dt \,=\, {1 \over K_1}(E_0(0)-E_0(+\infty))
   \,<\,+\infty~,
$$
hence $E_1(t)$ converges to $0$ as $t \to +\infty$. Similarly, combining 
\equ(E0), \equ(E2) and \equ(E3), we see that the positive functions 
$K_1 E_2(t)+K_4 E_0(t)$ and $K_1 E_3(t)+K K_4 E_0(t)$ are non-increasing
in time. Since $E_0(t)$ converges, the same is true for $E_2(t)$ and $E_3(t)$.
By \equ(E0) and \equ(E2), we have
$$
  \int_{0}^{+\infty}E_3(t)\,dt \,\le\, {1 \over K_1 K_3}
  \left(K_4(E_0(0)-E_0(+\infty))+ K_1(E_2(0)-E_2(+\infty))\right)\,<\,+\infty~,
$$
hence $E_3(t)$ converges to $0$ as $t \to \infty$. Since $\|(\partial_x f,
\partial_x g)\|_{X_s^2}^2 \le K_7(E_1 + E_3)$ for some $K_7 > 0$, 
\equ(gdecay) follows. \QED

We are now able to complete the proof of \clm(stab). 

\noindent {\bf Proof of \clm(stab).} Let $d_0 > 0$, $k_0 > 0$, $\epsilon > 0$
be as in the proof of \clm(B), and let $D \in [1-d_0,1+d_0]$, $k \in [0,k_0]$, 
$c \ge 2\sqrt{D}$. We set $\epsilon_0 = \epsilon/(2K_0)$, where $K_0$ is 
given by \clm(B). If $(f_0,g_0) \in Y_s^2$ satisfies $\|(f_0,g_0)\|_{Y_s^2} 
\le \epsilon_0$, then by \clm(local) there exists a time $t_1 > 0$ such that
\equ(fg) has a unique classical solution $(f, g) \in \CC^0([0,t_1],Y_s^2)
\cap \CC^1((0,t_1],Y_s^2)$ satisfying $(f(0),g(0)) = (f_0,g_0)$. By \clm(B), 
it follows that
$$
  \|(f(t),g(t))\|_{Y_s^2} \,\le\, K_0 \|(f_0,g_0)\|_{Y_s^2} \,\le\, 
  \epsilon/2~, \EQ(last)
$$
for all $t \in [0,t_1]$.  Indeed, assume that there exists $T \in (0,t_1]$ 
such that $\|(f(t), g(t))\|_{Y_s^2} < \epsilon$ for all $t \in [0, T)$ and 
$\|(f(T), g(T))\|_{Y_s^2} = \epsilon$. Then the hypothesis $\HH_\epsilon$ is 
satisfied on $[0,T]$, and \clm(B) implies that \equ(last) holds for 
$t \in [0,T]$, which is a contradiction. Therefore, $\|(f(t),g(t))\|_{Y_s^2} 
< \epsilon$ for all $t \in [0,t_1]$, and \equ(last) follows from \clm(B). This
shows that the solution $(f,g)$ of \equ(fg) in $Y_s^2$ satisfies \equ(last) 
whenever it exists. Using the Remark after \clm(local), we conclude that
$(f(t), g(t))$ exists for all $t \ge 0$ and satisfies \equ(last). This
proves \equ(bound), and \equ(limit) follows from \clm(H). 
This concludes the proof of \clm(stab). \QED  

\REFERENCES

\item{[AW]} D.G. Aronson and H.F. Weinberger: 
  {\pap Multidimensional Nonlinear Diffusion Arising in Population Genetics},
  Adv. in Math. {\bf 30} (1978), 33--76.   

\item{[BBDKL]} E. Ben-Jacob, H. Brand, G. Dee, L. Kramer and J.S. Langer:
  {\pap Pattern propagation in nonlinear dissipative systems},
  Physica {\bf 14D} (1985), 348--364. 

\item{[BX]} L. Berlyand and J. Xin:
  {\pap Large time asymptotics of solutions to a model combustion
  system with critical nonlinearity},
  Nonlinearity {\bf 8} (1995), 161--178.       

\item{[BN]} J. Billingham and D.J. Needham:
  {\pap The development of travelling waves in quadra\-tic and cubic
  autocatalysis with unequal diffusion rates. I. Permanent form 
  travelling waves}, Phil. Trans. R. Soc. Lond. A {\bf 334} (1991), 
  1--24. 

\item{[BK]} J. Bricmont and A. Kupiainen:
  {\pap Stability of Moving Fronts in the Ginzburg-Landau Equation}, 
  Comm. Math. Phys. {\bf 159} (1994), 287--318. 

\item{[BKX]} J. Bricmont, A. Kupiainen and J. Xin:
  {\pap Global large time self-similarity of a thermal-diffusive
  combustion system with critical nonlinearity},
  J. Diff Eqns. {\bf 130} (1996), 9--35.          

\item{[CX]} P. Collet and J. Xin:
  {\pap Global existence and large time asymptotic bounds of $\L^\infty$
  solutions of thermal diffusive combustion systems on $\real^n$},
  Ann. Scuola Norm. Sup. Pisa Cl. Sci., to appear. 

\item{[EW]} J.-P. Eckmann and C.E. Wayne:
  {\pap The Nonlinear Stability of Front Solutions for Parabolic Partial
  Differential Equations}, Comm. Math. Phys. {\bf 161} (1994), 323--334. 

\item{[FS]} E.B. Fabes and D.W. Stroock:
  {\pap A New Proof of Moser's Parabolic Harnack Inequality Using the
  Old Ideas of Nash}, Arch. Rat. Mech. Anal. {\bf 96} (1986), 327--338.

\item{[Fo1]} S. Focant: 
  {\pap Equations de r\'eaction-diffusion et \'etude de la stabilit\'e de
  fronts}, senior's thesis, Universit\'e Catholique de Louvain, 1993. 

\item{[Fo2]} S. Focant, in preparation. 

\item{[Ga1]} Th. Gallay: 
  {\pap Periodic patterns and travelling fronts for the Ginzburg-Landau 
  equation}, in IUTAM/ISIMM-Proceedings Nonlinear Waves in Hannover 1994, 
  World Scientific Press, 1995. 

\item{[Ga2]} Th. Gallay:
  {\pap Local Stability of Critical Fronts in Nonlinear Parabolic
  Partial Differential Equations}, Nonlinearity {\bf 7} (1994), 741--764. 

\item{[GR]} Th. Gallay and G. Raugel:
  {\pap Stability of travelling waves for a damped hyperbolic equation}, 
  to appear in ZAMP (1997). 

\item{[HY]} A. Haraux and A. Youkana:
  {\pap On a result of K. Masuda concerning reaction-diffusion equations},
  T\^ohoku Math. J. {\bf 40} (1988), 159--163. 

\item{[HPSS]} D. Horv\'ath, V. Petrov, S.K. Scott and K. Showalter: 
  {\pap Instabilites in propagating reaction-diffusion fronts},
  J. Chem. Phys. {\bf 98} (8) (1993), 6332--6343. 

\item{[Ki]} K. Kirchg\"assner:
  {\pap On the Nonlinear Dynamics of Travelling Fronts}, J. Diff. Eqns. 
  {\bf 96} (1992), 256--278. 

\item{[KR]} K. Kirchg\"assner and G. Raugel:                              
  {\pap Stability of Fronts for a KPP system : The Non-Critical             
  Case}, in {\bok Dynamics of nonlinear waves in dissipative systems:       
  reduction, bifurcation and stability}, G.~Dangelmayr, B.~Fiedler,         
  K.~Kirchg\"assner, A.~Mielke, Pitman Research Notes in Mathematics Series 
  {\bf 352}, Longman (1996).                                                

\item{[MP]} R.H Martin and M. Pierre:
  {\pap Nonlinear reaction-diffusion systems}, in {\bok Nonlinear  
  equations in the applied sciences}, W.F Ames and C. Rogers (Eds), 
  Academic Press, Boston (1992). 

\item{[Ma]} K. Masuda:
  {\pap On the global existence and asymptotic 
  behavior of solutions of reaction-diffusion equations}, Hokkaido 
  Math. J. {\bf 12} (1983), 360 --370. 

\item{[Pa]} A. Pazy:						  
  {\bok Semigroup of Linear Operators and Applications to Partial 
  Differential Equations}, Springer-Verlag (1983).                

\item{[Sa]} D.H. Sattinger:
  {\pap On the Stability of Waves of Nonlinear Parabolic Systems}, 
  Adv. Math. {\bf 22} (1976), 312--355. 

\item{[Tu]} A. Turing:
  {\pap The chemical basis of morphogenesis}, Phil. Trans. Roy. Soc. 
  Lond. {\bf B 237} (1952), 37--72. 

\item{[vS]} W. van Saarloos:
  {\pap Front propagation into unstable states. II. Linear versus 
  nonlinear marginal stability and rate of convergence} 
  Phys. Rev. A  {\bf 39} (12) (1989), 6367-6390. 

\end